# Intercellular coupling regulates the period of the segmentation clock

Leah Herrgen[1], Saúl Ares[2], Luis G. Morelli[1,2,3], Christian Schröter[1], Frank Jülicher[2], Andrew C. Oates[1]*

[1] Max Planck Institute for Molecular Cell Biology and Genetics, Pfotenhauerstr 108, 01307 Dresden, Germany.
[2] Max Planck Institute for the Physics of Complex Systems, Nöthnitzer Straße 38, 01187 Dresden, Germany.
[3] Departamento de Fisica, FCEyN, UBA, Pabellón I, Ciudad Universitaria, Buenos Aires, Argentina.
* Author for correspondence: oates@mpi-cbg.de

## *Summary*

**Background:** Coupled biological oscillators can tick with the same period. How this collective period is established is a key question in understanding biological clocks. We explore this question in the segmentation clock, a population of coupled cellular oscillators in the vertebrate embryo that sets the rhythm of somitogenesis, the morphological segmentation of the body axis. The oscillating cells of the zebrafish segmentation clock are thought to possess noisy autonomous periods, which are synchronized by intercellular coupling through the Delta-Notch pathway. Here we ask whether Delta-Notch coupling additionally influences the collective period of the segmentation clock.

**Results:** Using multiple embryo time-lapse microscopy, we show that disruption of Delta-Notch intercellular coupling increases the period of zebrafish somitogenesis. Embryonic segment length and the spatial wavelength of oscillating gene expression also increase correspondingly, indicating an increase in the segmentation clock's period. Using a theory based on phase oscillators in which the collective period self-organizes due to time delays in coupling, we estimate the cell-autonomous period and the delay in coupling from our data. Further supporting the role of coupling delays in the clock, we predict and experimentally confirm an instability resulting from decreased coupling delay time.

**Conclusions:** Synchronization of cells by Delta-Notch coupling regulates the collective period of the segmentation clock. Our identification of the first segmentation clock period mutants is a critical step towards a molecular understanding of temporal control in this system. We propose that collective control of period via delayed coupling may be a general feature of biological clocks.

## *Highlights*

- The period of the vertebrate segmentation clock is regulated by Delta-Notch signaling.
- Communication between oscillating cells in the embryo regulates their collective period.
- Delays in the communication time are likely critical for the alteration of the period.

*Introduction*

Many biological clocks are composed of coupled autonomous oscillators [1]. Generally, a population of oscillators with differing autonomous periods can synchronize through coupling, choosing a collective period that is the average of the autonomous periods of individual oscillators [2] (Fig.1A). Examples are coupled chemical systems [3], chirping crickets [4], and flashing fireflies [5]. However, if a significant delay occurs in the coupling, the dynamical effects of the delays may result in novel and complex synchronization phenomena [6, 7]. In particular, the collective period can differ significantly from the average period of autonomous oscillations (Fig. 1A). While this has been studied in mathematical models, and in engineered systems of coupled lasers [8], examples of this scenario have not been identified in biological systems.

The zebrafish segmentation clock may provide an example of such a system. This clock is a population of coupled cellular genetic oscillators in the embryo that drives the sequential subdivision of the presomitic mesoderm (PSM) into multi-cellular blocks termed somites, with a period of approximately 25 min [9-14]. Upon dissociation, individual PSM cells behave as noisy autonomous oscillators with a range of different periods [11], but within the PSM, neighboring cellular oscillators are synchronized by coupling through the Delta-Notch signal transduction pathway [10, 13, 15, 16] (Fig. 2A). Reduction of Delta-Notch coupling results in the gradual decay of segmentation clock synchrony, eventually causing somite boundary defects at a position along the embryonic axis that depends on the remaining strength of the coupling [10, 13, 15]. Given that delays in the coupling on the order of the period are expected from synthesis, trafficking and transduction of Delta coupling signals [17, 18], it is possible that Delta-Notch coupling causes a collective period in the zebrafish segmentation clock that differs from the average of the autonomous oscillators, in agreement with numerical simulation of gene network models of the zebrafish segmentation clock for several cells [9, 19].

Together with these temporal properties, the segmentation clock also has a spatial aspect. Oscillating gene expression waves propagate through the PSM, forming striped patterns [12]. These patterns of gene expression are consistent with a gradual slowing of autonomous oscillator frequency as cells approach an arrest front in the anterior PSM, a situation that can be described by oscillators with a position-dependent frequency along the PSM [12, 17, 20-25]. We recently introduced a general framework building on previous efforts [13, 17, 21, 22] that combines the spatial and temporal aspects of the segmentation clock and adds delayed coupling between neighboring oscillators [20] (Fig. 1B). This delayed coupling theory describes the collective spatiotemporal behavior of the tissue arising from properties of individual cells, which are modeled as phase oscillators. Because the delayed coupling theory has a relatively small number of parameters, most of which can be measured from the developing embryo, it is well suited to fit experimental data. The oscillating gene expression patterns described by the theory are in good quantitative agreement with experimental data from wildtype embryos [17, 20]. Importantly, this approach permits the quantitative experimental analysis of changes to the pattern's wavelength induced by altered collective period or frequency profile [20].

According to this theoretical work, if significant time delays exist in Delta-Notch coupling, then reduction of this coupling will change the collective period of the segmentation clock. This is expected to give rise to three defining phenotypes in the embryo: (i) an altered somitogenesis period [9, 19, 20], (ii) a correspondingly altered segment length due to the change in period [12, 17, 20-22, 26-28], and (iii) corresponding changes to the expression pattern wavelength of oscillating genes in the PSM [20] (Fig. 1C, Movie S1). In addition, the delayed coupling theory predicts that there are values of the coupling delay for which a dynamic instability occurs [20], destabilizing the spatial patterns of gene expression. In this work we experimentally confirm these predictions and use the delayed coupling theory to estimate from our data the period of the autonomous oscillations, the coupling strength, and the time delay in the coupling. Together, these findings show that the segmentation clock's collective period is regulated by the Delta-Notch pathway, and support a role for delayed coupling in setting the collective period of biological clocks.

*Results*
**Somitogenesis period is increased by reduction of Delta-Notch coupling**
An altered segmentation clock period is expected to alter the period of somitogenesis. To test if reduced Delta-Notch coupling alters the period of somitogenesis, we analyzed conditions affecting this signaling pathway. These were mutants for the Delta ligands *after eight (aei/deltaD)* and *beamter (bea/deltaC)*, the Notch receptor *deadly seven (des/notch1a)*, and the E3 ubiquitin ligase *mind bomb (mib)* involved in Delta trafficking, as well as embryos treated with DAPT, a γ-secretase inhibitor that attenuates Notch receptor function [10, 13] (Fig. 2A, Experimental procedures). Somitogenesis period was measured in live embryos using our high-precision time-lapse microscopy protocol [14, 29]. Somitogenesis period was increased to 119 ± 2% (mean ± 95% confidence interval, CI) in *mib* (Fig. 2B-E; Movies S2, S3), to 123 ± 4% in *aei/deltaD*, and to 107±3% in *des/notch1a* embryos (Fig. 2E) as compared to their wildtype siblings. The increase was constant until intact somite boundaries no longer formed due to loss of synchrony (Fig. 2C) [13, 30]. Blocking Notch signaling with saturating DAPT concentrations (≥ 40 μM) [13] increased somitogenesis period to 118 ± 1% of control (Fig. 2E), indicating that the smaller change in *des/notch1a* is likely due to Notch receptor redundancy [13].

The morphology of the forming anterior somite boundaries in *bea/deltaC* embryos is less regular than with DAPT treatment and in the other Delta-Notch mutants [31, 50], and this prevented precise measurement of somitogenesis period in our assay. Although *bea/deltaC* and DAPT treatment yield the same desynchronization phenotype, indicating an equivalent quantitative contribution to coupling within the clock [55, 56, 13], the combined observations on anterior somite boundary morphology suggest an additional, γ-secretase inhibitor insensitive role for *bea/deltaC* in somite boundary formation downstream of the segmentation clock.

We detected no change in general developmental rate and tissue differentiation (Fig. S1), or in embryonic axial elongation rate (Fig. S2) in the Delta-Notch mutant or DAPT-treated embryos, consistent with previous studies [30, 31]. This argues against general slowing of development as an explanation for increased somitogenesis period. Fibroblast Growth Factor, Wnt, and Retinoic Acid signaling are required for aspects of vertebrate somitogenesis, but are not implicated in coupling, as reviewed in [32]. We measured somitogenesis period in *acerebellar (ace/fgf8)*, *masterblind (mbl/axin1)* and *no fin (nof/raldh2)* mutants, which affect the FGF, Wnt, and Retinoic Acid pathways respectively, and detected no change (Fig. 2E). While not excluding roles for these pathways in period setting, these results indicate that increased somitogenesis period is not a general consequence of defective intercellular PSM signaling pathways. We conclude that Delta-Notch coupling regulates somitogenesis period.

**Segment length is increased by reduction of Delta-Notch coupling**

Segment length in the elongating embryo is thought to be determined by interaction of the segmentation clock and a posteriorly-moving wavefront of rapid cell change that arrests oscillations at the anterior end of the PSM [26, 28, 32]. In this Clock and Wavefront model, if the segmentation clock's rhythm is translated directly into spatial periodicity, then the resulting segment length is given by $S = vT_C$, with segment length $S$, arrest front velocity $v$, and the segmentation clock's collective period $T_C$ [12, 17, 20-22, 26-28]. Embryonic axial elongation rate and the posterior border of the *mespb* gene expression domain, which we use to define the arrest front location within the PSM [33], were not significantly different between control and Delta-Notch impaired embryos throughout the time of interest (Fig. S3, Movies S4, S5), showing that arrest front velocity was unchanged. Consequently, with an increased segmentation clock period, we expect longer segments. Anterior-posterior length of somites 2-5 was increased in live *mib*, *aei/deltaD* and DAPT-treated embryos versus wildtype siblings (Fig. 3A, B), but in *des/notch1a* embryos, which had the smallest period change, we detected no alteration of somite length (Fig. 3B). Assaying the distance between *mespb* expression stripes, which we use to mark segment length at the arrest front before somite formation [33], revealed an increase in segment length in *mib*, *aei/deltaD*, *des/notch1a* and DAPT-treated embryos (Fig. 3C, D). We conclude that Delta-Notch coupling has an effect on segment length.

**Spatial patterns of oscillating cells are altered in *mind bomb* mutants**

If the changes we observed in somitogenesis period are produced by an altered collective period $T_C$ of the segmentation clock, then we should find corresponding changes to the gene expression patterns in the PSM. These changes can be quantified [20] and should provide an independent, gene expression-based estimate for the period change measured in our time-lapse experiments. To test this prediction, we quantified stripe wavelength in *mib* mutant embryos, finding a systematic increase in wavelength for a given position as compared to wildtype siblings (Fig. 4A,B). Fitting the delayed coupling theory to the data (Fig. 4B; Experimental procedures) indicated an increase in normalized segment length *s* of 116±1% in *mib* embryos while leaving the frequency profile unchanged (Fig. 4C). Since arrest front velocity is not altered in the mutant, $S = vT_C$ indicates a change to collective period similar to that obtained from somitogenesis period measurement (119±2%, Fig. 2).

In summary, we have shown that conditions reducing Delta-Notch signaling lead to corresponding changes in (i) somitogenesis period and (ii) segment length, without changes to arrest front velocity. In addition, (iii) we find changes to oscillating gene expression patterns in the PSM consistent with changes in collective period (Table 1). Combined, our findings indicate a change in the segmentation clock's collective period due to reduced Delta-Notch coupling. This is as expected if significant delays exist in the coupling. The magnitude of these changes, ~10-20% of the wildtype period, is similar to those resulting from single gene mutations in the circadian clock [34-37].

**Estimation of autonomous period and coupling delay from the data**

In order to understand these collective period changes in terms of the strength and delay in coupling between PSM cells, we again use the delayed coupling theory [20]. The collective period $T_C$ is related to the period $T_A$ of uncoupled autonomous oscillators in the posterior PSM, a positive coupling strength $\varepsilon$ and coupling delay $\tau$ by:

$$\frac{2\pi}{T_C} = \frac{2\pi}{T_A} - \varepsilon \sin\left(\frac{2\pi}{T_C}\tau\right). \qquad (1)$$

Note that the coupling strength $\varepsilon$ and coupling delay $\tau$ are effective parameters of a tissue-level theory. Therefore, their values are not necessarily related in simple ways to molecular quantities such as the number of molecules involved in coupling or duration of a molecular signaling event. Nevertheless, their values depend on underlying molecular processes, and changes to these processes cause corresponding changes to the effective parameters (Experimental Procedures). The effects of delayed coupling on collective period in Eq. (1) are illustrated with the following *Gedankenexperiment*: If two uncoupled oscillators cycling at the same pace with equal phases are suddenly coupled with a delay, they will receive information about the phase of the other oscillator from an earlier time point. The oscillators will change period as they try to synchronize to the apparent (delayed) phase of their neighbor. Depending on the relative value of the delay and the autonomous period, this could result in slowing or speeding of the oscillators. As this process occurs across a population, a collective period self-organizes.

Eq. (1) predicts a continuous variation of collective period, and therefore somitogenesis period, with coupling strength. We tested this prediction by determining the response of somitogenesis period to varying DAPT concentrations and found a variation up to saturation consistent with a smooth change of the collective period (Fig. 5A).

For a given autonomous period $T_A$, the collective period $T_C$ described by Eq. (1) depends on coupling strength $\varepsilon$ and coupling delay $\tau$ in a complex way (Fig. 5B), analogous to that described in models of gene regulatory networks with delay in the coupling [38]. As the time delay is varied, branches of stable synchronized oscillations are separated by unstable branches where synchrony is lost (Fig. 5B). To locate the situation corresponding to the wildtype zebrafish segmentation clock in this diagram, parameters $\varepsilon$, and $\tau$ were estimated by fitting the delayed coupling theory to our experimental data (Fig. 5A, Experimental Procedures). For this purpose, we assume that saturating DAPT completely blocks coupling [13] and that a loss of coupling does not affect $T_A$ over relevant time scales [13, 15]. From our timelapse experiments, the measured value of $T_A \approx 1.18 T_C$ (Fig. 2E), which for the zebrafish embryo at 28°C [14] yields $T_A = 28 \pm 1$ min. This value is consistent with independent estimates for $T_A$ from genetic network models [9, 19]. To account for the measured change in period, coupling strength must be at least 0.043 min$^{-1}$ (Eq. (1)) and from the fit, we estimate coupling strength $\varepsilon = 0.07 \pm 0.04$ min$^{-1}$.

Multiple values of the coupling delay $\tau$ are consistent with the wildtype collective period $T_C = 23.5$ min (Fig. 5B, dotted line). We use the only value of $\tau$ for which the wave pattern is both stable and unique for the determined values of $T_A$ and $\varepsilon$, which is found on branch 2 (Fig. 5B), and so estimate coupling delay $\tau = 21 \pm 2$ min, corresponding to $0.9T_C$.

The various experimental perturbations of the Delta-Notch pathway can be consistently mapped to distinct points in a diagram relating collective period and coupling delay (Fig. 5C). This map uses the assumption in accordance with previous work [9] that most of the delay in Delta-Notch signaling in the segmentation clock arises from the synthesis and trafficking of Delta molecules in the signal-sending cell [18], and that signal transduction via the cleavage of the Notch intracellular domain is on a shorter time scale [39]. Thus, all mutants and treatments studied here affect the coupling strength, which can be estimated from the onset of segmental defects caused by the gradual desynchronization of the segmentation clock's cells [10, 13, 15, 16], but only *aei/deltaD* and *mib* affect the coupling delay in addition. The *mib* mutant, with reduced endocytic trafficking of Delta in the signal-sending cell [40], shows a strong collective period change despite exhibiting the weakest somite boundary disruption phenotype of any mutant (Table 1), suggesting that coupling delay is strongly increased but coupling strength is only mildly affected. Heterozygote *mib* embryos show an increased period without any somite defects (Fig. S4), suggesting that the coupling delay depends sensitively on the level of Mib protein.

**Mind bomb over-expression drives the system into instability**

A key prediction of the delayed coupling theory is the existence of instabilities for certain values of the coupling delay. In particular, the instability that separates branch 1 from branch 2 should be reached by reducing the coupling delay from the wildtype value of 21 min to about 16 min (Fig. 5B,C). Given that reduced levels of Mib increase coupling delay, we reasoned that elevated levels of Mib might offer a way to shorten the delays and thereby experimentally test the existence of the predicted instability. We first performed numerical simulations of the PSM using the delayed coupling theory with wildtype parameters determined in this work. We found an excellent agreement with the spatial organization of *dlc* expression in the wildtype embryo that can be quantified using an autocorrelation function (Fig. 6A-C; Experimental procedures). We next repeated the simulation with reduced coupling delay. We found that as the system approaches the instability, the normal striped gene expression patterns develop a characteristic disrupted pattern with a range of spatial wavelengths resulting in a flattened average autocorrelation function (Fig. 6D,E; Fig. S5). To test this prediction in vivo, we injected *mib* mRNA into the embryo and assayed the resulting spatial features of cyclic *dlc* expression in the PSM. We found that elevated Mib levels resulted in a disruption of somitogenesis in otherwise normally developing embryos accompanied by a disrupted pattern of *dlc* expression with an autocorrelation function in quantitative agreement with simulations of reduced delay (Fig. 6D,E; Fig. S6). These patterns are distinct from those observed in the loss of coupling mutants *aei/deltaD*, *des/notch1a* and *mib* (Fig. S7). These data provide evidence of the predicted delay-dependent dynamic instability, and give further support to the existence of delays in Delta-Notch coupling in the zebrafish segmentation clock.

*Discussion*

Our work in this paper was motivated by theoretical results suggesting that systems of oscillators with time delays in the coupling could tick with a collective period different from the period of the autonomous oscillators (Fig. 1). Coupling by Delta-Notch signaling in the zebrafish segmentation clock was an attractive candidate system to observe such effects because the clock's period is of the same order as expected signaling delays. Here we describe that a reduction in Delta-Notch coupling increases somitogenesis period (Fig. 2), produces longer segments (Fig. 3), and lengthens the wavelength of oscillating gene expression stripes (Fig. 4) in an otherwise normally developing embryo. For this study, the use of multiple-embryo time-lapse imaging [14, 29] was critical to sensitively and precisely measure the timing of somitogenesis and growth in live embryos under controlled conditions. Combined, these experimental results are compelling evidence of a role for Delta-Notch signaling in controlling the period of the segmentation clock.

We analyzed the role of the Delta-Notch intercellular signaling system in regulating the period using the delayed coupling theory [20], which describes the segmentation clock in a simplified way at a cellular and tissue level as a spatially distributed population of phase oscillators with time delays in the coupling. According to theory, collective period is modulated through perturbation of coupling if there are time delays in the coupling [6, 7, 9, 19, 20]. The existence of coupling delays in the segmentation clock is supported by the fit of the delayed coupling theory to the data from the change in collective period with coupling strength (Fig. 5), as well as the successful numerical simulation of the oscillating gene expression patterns of the wildtype PSM (Fig. 6) using the values of autonomous period, coupling strength, and coupling delay obtained from the fit. Further support for a critical role of coupling delay in the system comes from the experimental observation of a predicted instability consistent with shorter delays (Fig. 6). This comparison of spatial patterns of cyclic gene expression in experiment and simulation using an autocorrelation function introduces a novel method for the analysis of perturbation to the segmentation clock. Together, the fit of theory and experimental data in this work gives strong support to the existence and period-setting function of delays in Delta-Notch coupling in the segmentation clock.

Phase oscillators have been shown to correctly capture the dynamics of gene regulatory network models of the segmentation clock [23]. Building such models is one of the important goals of the field [9, 19, 23, 24, 41-43], but the experimental measurement of the many rate parameters of these models is currently difficult in vivo. In contrast, most parameters of the delayed coupling theory, such as collective period, arrest front velocity, segment length, and frequency profile, can be measured from the embryo directly or have been estimated previously [14, 20], leaving autonomous period, coupling strength, and coupling delay to be obtained in this work from the fit of the theory to the new data (Fig. 5). The main use of the values we estimate for the effective parameters coupling strength and coupling delay is in the quantitative comparison of experimental situations. Future work may connect these parameter values to measurable events described in detailed molecular models. In contrast, the autonomous period is a simple biological parameter that can be directly tested using methods with cellular resolution of the dynamics.

Our discovery of segmentation period mutants in an intercellular signaling system provides insight into the fundamentally multi-cellular organization of the segmentation clock. While it was previously recognized that coupling in the zebrafish segmentation clock synchronizes the phases of the oscillating cells of the PSM [10, 13, 15, 16], the results here indicate that the time cost of coupling using macromolecules causes an additional, novel effect – a self-organized regulation of the collective period. This phenomenon is likely not restricted to the segmentation clock, since any system where oscillators are coupled with delays of the order of the period should show similar effects. One biological candidate system is the circadian clock, in which cellular rhythms are synchronized by VIP signaling, and where loss of this coupling alters the period [44, 45]. A similar situation may occur in other multi-cellular systems with coordinated oscillations such as the neuroepithelium and hair follicles [46, 47]. The role of collective effects in controlling the period of other biological clocks now awaits investigation.

## *Experimental Procedures*

In this section we provide a brief outline of the methods that we introduce and use in this work. This material, plus a more comprehensive description and extensive technical details is provided in corresponding sections in the Supplemental Information. Embryology, microscopy, and molecular biology are covered in section 2.1 and theoretical methods and parameter fitting are covered in section 2.2.

### Embryology, microscopy, and molecular biology

Zebrafish embryos were obtained using standard procedures [48]. Mutant alleles were *aei* $^{tr233}$ [49], *bea* $^{tm98}$ [50], *des* $^{p37a}$ [51], *mib* $^{ta52b}$ [40], *ace* $^{ti282a}$ [52], *mbl* $^{tm013}$ [53] and *nof* $^{u11}$ [54], and DAPT treatment was according to [13]. Timelapse recordings and estimation of somitogenesis period were according to [14, 29], and in situ hybridization according to [55]. Statistical significance in somitogenesis period, somite and segment length, axial elongation rate, and position of the arrest front was assessed with Student's t-test, two-sided, unequal variance. In vitro synthesis of *mib* and *GFP* mRNA was carried out as previously described [56], quantified using a Nanodrop1000 spectrophotometer (Thermo Scientific), and delivered in varying amounts to one-cell stage embryos using our quantitative injection protocol [13]. After injection, embryos were left to develop until they reached the 3-4 somite stage when they were fixed for in situ hybridization.

### Theoretical methods and parameter fitting

*Overview of the delayed coupling theory*

The delayed coupling theory (DCT) describes the PSM as an array of coupled phase oscillators, and has been previously introduced in detail [20]. The state of oscillator $i$ at time $t$ is described by a phase variable $\theta_i(t)$ whose dynamics is given by

$$\frac{d\theta_i(t)}{dt} = \omega_i(t) + \frac{\varepsilon_i(t)}{N_i}\sum_k \sin[\theta_k(t-\tau) - \theta_i(t)] + \eta\zeta_i(t) \ . \tag{2}$$

The frequency profile $\omega_i(t)$ describes the slowing down of oscillators as they get closer to the arrest front, where oscillations stop. Oscillators are coupled to their neighbors with a coupling strength $\varepsilon_i(t)$, and a delay $\tau$ accounting for the finite time introduced by synthesis and trafficking of ligands in the cells sending the signal. Oscillator $i$ is coupled to its $N_i$ nearest neighbors, labeled by $k$. The zero average un-correlated random term $\zeta_i$ represents different noise sources with total noise strength $\eta$.

*Parameter estimation from cyclic stripe wavelength fitting*
Cyclic gene expression patterns are related to the frequency profile in the DCT. To fit these patterns, it is convenient to use the continuum approximation to Eq. (2) in the reference frame co-moving with the PSM [20]. We assume that the intrinsic frequency at a distance $x$ to the arrest front is $\omega(x) = \omega_A(1-e^{-x/\sigma})/(1-e^{-L/\sigma})$, where $x$ is a continuous variable, $\omega_A = 2\pi/T_A$ is the frequency of the individual oscillators in the posterior PSM. $\sigma$ is the decay length of the frequency profile, measuring the characteristic distance over which the frequency decreases from high to low values, and $L$ is the distance between the posterior end of the notochord and the arrest front of the oscillations. The continuum description relates the position $x$ of the centre of a wave of gene expression with its wavelength $\lambda$ (Fig. 4A) through the expression [20]:

$$x = \sigma \ln\left[\frac{2\sigma\sinh(\lambda/2\sigma)}{S(1-e^{-L/\sigma}) + \lambda e^{-L/\sigma}}\right], \tag{3}$$

where $S$ is the arrested segment length. Note that the patterns of expression described by Eq. (3) are not sensitive to spatial variation in coupling strength or delay. Fits of the theory to our *deltaC* stripe wavelength measurements were done using a total least square method, and error bars for $S$ and $\sigma$ were calculated using a bootstrap method. To allow an intuitive visualization of the results and comparison with potential results in different species, in Figures 4B,C we show the results normalized by the length $L$, for which we defined the normalized parameters $\rho = \sigma/L$ and $s = S/L$.

*Fit of somitogenesis period vs. DAPT treatment concentration*
To relate the coupling strength $\varepsilon$ in Eq. (1) to the concentration $n$ of the DAPT treatment, we first assume [13] that DAPT suppresses Notch signaling following Michaelis-Menten kinetics, [Notch]~$1/(1+n/n_0)$, where $n_0$ is the DAPT concentration that halves Notch signaling. Assuming a linear relation of coupling strength to the level of Notch signaling, and that saturating DAPT completely blocks coupling [13] then yields $\varepsilon(n) = B/(1+n/n_0)$, where $B$ is a constant related to the wildtype level of Notch receptor expression. Following previous work, we assume that DAPT does not affect $T_A$ over relevant time scales [13, 15]. Starting from Eq. (1) a relation between period of oscillation and the level of DAPT treatment can be derived. Including a factor 100 to express the result as a percentage and assuming that the coupling delay lies on the second branch (Fig. 5B) this relation is:

$$\frac{T(n)}{T_u}\% = 100\frac{nT_A/T_u + Q}{n + Q}, \quad (4)$$

$$Q = n_0(1 + BT_A). \quad (5)$$

$T_u = T(0)$ is the period of untreated wildtype embryos. The parameters $n_0$ and $B$ appear only through the combination $Q$ defined in Eq. (5). From the fit of Eq. (4) to the data in Fig. 5A we obtain $Q = 7.5 \pm 3.5$ µM. Using this value of $Q$ we can estimate the coupling strength $\varepsilon = \varepsilon(0) = B = 0.07 \pm 0.04$ min$^{-1}$ and the coupling delay $\tau = 21 \pm 2$ min.

*Numerical simulations of the delayed coupling theory*

For the simulations in this work we used a hexagonal two-dimensional lattice that has the same average shape and size as the experimental conditions. We solved the DCT equations using parameters that we estimated in this work for the wildtype zebrafish. We measured a cell size of 7 pixels, and used this as the size of single cells in the simulated patterns. Other parameters are the decay length of the frequency profile [20] $\sigma = 27$ cell diameters, the autonomous frequency of oscillators at the posterior boundary $\omega_A = 0.2205$ min$^{-1}$, the coupling strength $\varepsilon = 0.07$ min$^{-1}$, the coupling delay $\tau = 20.75$ min, and the arrest front velocity $v = 0.249$ cell diameters/min. Fluctuations were introduced with a noise term of strength $\eta = 0.02$ min$^{-1}$.

*Autocorrelation function*

In this work we introduce a spatial autocorrelation function to measure disorder in gene expression patterns under different experimental conditions. We compute the autocorrelation function in a box from the anterior-most region of the PSM, where the most striking features of the expression patterns of wildtype, mutant, and also the Mib over-expression experiments are observed (Figure 6A). We define the autocorrelation function of the observed fluorescence intensity $I(x)$ as

$$C(\delta) = \langle I(x)I(x+\delta)\rangle \quad (6)$$

where the position $x$ is measured from the anterior border of the box along the antero-posterior axial direction, and the brackets denote an average over space along the same axial direction followed by an average across the lateral direction. The autocorrelation is a function of the distance $\delta$ between two points in the pattern. Peaks of this function occur at distances where the intensity of the pattern is on average similar, as for example in consecutive stripes or interstripes of the cyclic gene expression patterns.

*Acknowledgements*

The authors would like to thank the MPI-CBG fish, light microscopy, and sequencing facilities, the Dresden fish community, members of the Oates and Jülicher groups for discussion, and L. Rhode, I. Riedel-Kruse, J. Howard, C.-P. Heisenberg, J. Pecreaux for critical comments on the manuscript. This work was funded by the Max Planck Society and by the European Research Council under the European Communities Seventh Framework Programme (FP7/2007-2013) / ERC Grant no. 207634.

*References*

1. Winfree, A.T. (1967). Biological rhythms and the behavior of populations of coupled oscillators. J Theor Biol *16*, 15-42.
2. Kuramoto, Y. (1984). Chemical Oscillations, Waves and Turbulence, (Berlin: Springer Verlag).
3. Taylor, A.F., Tinsley, M.R., Wang, F., Huang, Z., and Showalter, K. (2009). Dynamical quorum sensing and synchronization in large populations of chemical oscillators. Science *323*, 614-617.
4. Walker, T.J. (1969). Acoustic Synchrony: Two Mechanisms in the Snowy Tree Cricket. Science *166*, 891-894.
5. Buck, J. (1988). Synchronous rhythmic flashing of fireflies. II. Q Rev Biol *63*, 265-289.
6. Schuster, and Wagner (1989). Mutual entrainment of two limit cycle oscillators with time delayed coupling. Prog Theor Phys *81*, 939-945.
7. Yeung, S., and Strogatz, S. (1999). Time Delay in the Kuramoto Model of Coupled Oscillators. Phys. Rev. Lett. *82*, 648-651.
8. Wunsche, H.J., Bauer, S., Kreissl, J., Ushakov, O., Korneyev, N., Henneberger, F., Wille, E., Erzgraber, H., Peil, M., Elsasser, W., et al. (2005). Synchronization of delay-coupled oscillators: a study of semiconductor lasers. Phys Rev Lett *94*, 163901.
9. Lewis, J. (2003). Autoinhibition with transcriptional delay: a simple mechanism for the zebrafish somitogenesis oscillator. Curr Biol *13*, 1398-1408.
10. Horikawa, K., Ishimatsu, K., Yoshimoto, E., Kondo, S., and Takeda, H. (2006). Noise-resistant and synchronized oscillation of the segmentation clock. Nature *441*, 719-723.
11. Masamizu, Y., Ohtsuka, T., Takashima, Y., Nagahara, H., Takenaka, Y., Yoshikawa, K., Okamura, H., and Kageyama, R. (2006). Real-time imaging of the somite segmentation clock: revelation of unstable oscillators in the individual presomitic mesoderm cells. Proc Natl Acad Sci U S A *103*, 1313-1318.
12. Palmeirim, I., Henrique, D., Ish-Horowicz, D., and Pourquie, O. (1997). Avian hairy gene expression identifies a molecular clock linked to vertebrate segmentation and somitogenesis. Cell *91*, 639-648.
13. Riedel-Kruse, I.H., Muller, C., and Oates, A.C. (2007). Synchrony dynamics during initiation, failure, and rescue of the segmentation clock. Science *317*, 1911-1915.
14. Schroter, C., Herrgen, L., Cardona, A., Brouhard, G.J., Feldman, B., and Oates, A.C. (2008). Dynamics of zebrafish somitogenesis. Dev Dyn *237*, 545-553.
15. Jiang, Y.J., Aerne, B.L., Smithers, L., Haddon, C., Ish-Horowicz, D., and Lewis, J. (2000). Notch signalling and the synchronization of the somite segmentation clock. Nature *408*, 475-479.
16. Ozbudak, E.M., and Lewis, J. (2008). Notch signalling synchronizes the zebrafish segmentation clock but is not needed to create somite boundaries. PLoS Genet *4*, e15.
17. Giudicelli, F., Ozbudak, E.M., Wright, G.J., and Lewis, J. (2007). Setting the tempo in development: an investigation of the zebrafish somite clock mechanism. PLoS Biol *5*, e150.
18. Heuss, S.F., Ndiaye-Lobry, D., Six, E.M., Israel, A., and Logeat, F. (2008). The intracellular region of Notch ligands Dll1 and Dll3 regulates their trafficking and signaling activity. Proc Natl Acad Sci U S A *105*, 11212-11217.
19. Leier, A., Marquez-Lago, T.T., and Burrage, K. (2008). Generalized binomial tau-leap method for biochemical kinetics incorporating both delay and intrinsic noise. J Chem Phys *128*, 205107.
20. Morelli, L., Ares, S., Herrgen, L., Schroeter, C., Juelicher, F., and Oates, A. (2009). Delayed Coupling Theory of Vertebrate Segmentation. HFSP J *1*, 85.
21. Kaern, M., Menzinger, M., and Hunding, A. (2000). Segmentation and somitogenesis derived from phase dynamics in growing oscillatory media. J Theor Biol *207*, 473-493.
22. Jaeger, J., and Goodwin, B.C. (2001). A cellular oscillator model for periodic pattern formation. J Theor Biol *213*, 171-181.
23. Uriu, K., Morishita, Y., and Iwasa, Y. (2009). Traveling wave formation in vertebrate segmentation. J Theor Biol *257*, 385-396.

24. Cinquin, O. (2007). Repressor dimerization in the zebrafish somitogenesis clock. PLoS Comput Biol *3*, e32.
25. Tiedemann, H.B., Schneltzer, E., Zeiser, S., Rubio-Aliaga, I., Wurst, W., Beckers, J., Przemeck, G.K., and Hrabe de Angelis, M. (2007). Cell-based simulation of dynamic expression patterns in the presomitic mesoderm. J Theor Biol *248*, 120-129.
26. Cooke, J., and Zeeman, E.C. (1976). A clock and wavefront model for control of the number of repeated structures during animal morphogenesis. J Theor Biol *58*, 455-476.
27. Gomez, C., Ozbudak, E.M., Wunderlich, J., Baumann, D., Lewis, J., and Pourquie, O. (2008). Control of segment number in vertebrate embryos. Nature *454*, 335-339.
28. Cooke, J. (1981). The problem of periodic patterns in embryos. Phil. Trans. R. Soc. Lond. B *295*, 509 - 524.
29. Herrgen, L., Schroter, C., Bajard, L., and Oates, A.C. (2009). Multiple embryo time-lapse imaging of zebrafish development. Methods Mol Biol *546*, 243-254.
30. Jiang, Y.J., Brand, M., Heisenberg, C.P., Beuchle, D., Furutani-Seiki, M., Kelsh, R.N., Warga, R.M., Granato, M., Haffter, P., Hammerschmidt, M., et al. (1996). Mutations affecting neurogenesis and brain morphology in the zebrafish, Danio rerio. Development *123*, 205-216.
31. van Eeden, F.J., Granato, M., Schach, U., Brand, M., Furutani-Seiki, M., Haffter, P., Hammerschmidt, M., Heisenberg, C.P., Jiang, Y.J., Kane, D.A., et al. (1996). Mutations affecting somite formation and patterning in the zebrafish, Danio rerio. Development *123*, 153-164.
32. Dequeant, M.L., and Pourquie, O. (2008). Segmental patterning of the vertebrate embryonic axis. Nat Rev Genet *9*, 370-382.
33. Sawada, A., Fritz, A., Jiang, Y.J., Yamamoto, A., Yamasu, K., Kuroiwa, A., Saga, Y., and Takeda, H. (2000). Zebrafish Mesp family genes, mesp-a and mesp-b are segmentally expressed in the presomitic mesoderm, and Mesp-b confers the anterior identity to the developing somites. Development *127*, 1691-1702.
34. Konopka, R.J., and Benzer, S. (1971). Clock mutants of Drosophila melanogaster. Proc Natl Acad Sci U S A *68*, 2112-2116.
35. Bruce, V.G. (1972). Mutants of the biological clock in Chlamydomonas reinhardi. Genetics *70*, 537-548.
36. Feldman, J.F., and Hoyle, M.N. (1973). Isolation of circadian clock mutants of Neurospora crassa. Genetics *75*, 605-613.
37. Vitaterna, M.H., King, D.P., Chang, A.M., Kornhauser, J.M., Lowrey, P.L., McDonald, J.D., Dove, W.F., Pinto, L.H., Turek, F.W., and Takahashi, J.S. (1994). Mutagenesis and mapping of a mouse gene, Clock, essential for circadian behavior. Science *264*, 719-725.
38. Momiji, H., and Monk, N.A. (2009). Oscillatory Notch-pathway activity in a delay model of neuronal differentiation. Phys Rev E Stat Nonlin Soft Matter Phys *80*, 021930.
39. Shimizu, K., Chiba, S., Hosoya, N., Kumano, K., Saito, T., Kurokawa, M., Kanda, Y., Hamada, Y., and Hirai, H. (2000). Binding of Delta1, Jagged1, and Jagged2 to Notch2 rapidly induces cleavage, nuclear translocation, and hyperphosphorylation of Notch2. Mol Cell Biol *20*, 6913-6922.
40. Itoh, M., Kim, C.H., Palardy, G., Oda, T., Jiang, Y.J., Maust, D., Yeo, S.Y., Lorick, K., Wright, G.J., Ariza-McNaughton, L., et al. (2003). Mind bomb is a ubiquitin ligase that is essential for efficient activation of Notch signaling by Delta. Dev Cell *4*, 67-82.
41. Goldbeter, A., and Pourquie, O. (2008). Modeling the segmentation clock as a network of coupled oscillations in the Notch, Wnt and FGF signaling pathways. J Theor Biol *252*, 574-585.
42. Gonzalez, A., and Kageyama, R. (2009). Hopf bifurcation in the presomitic mesoderm during the mouse segmentation. J Theor Biol *259*, 176-189.
43. Uriu, K., Morishita, Y., and Iwasa, Y. (2009). Synchronized oscillation of the segmentation clock gene in vertebrate development. J Math Biol.
44. Aton, S.J., Colwell, C.S., Harmar, A.J., Waschek, J., and Herzog, E.D. (2005). Vasoactive intestinal polypeptide mediates circadian rhythmicity and synchrony in mammalian clock neurons. Nat Neurosci *8*, 476-483.
45. Renn, S.C., Park, J.H., Rosbash, M., Hall, J.C., and Taghert, P.H. (1999). A pdf neuropeptide gene mutation and ablation of PDF neurons each cause severe abnormalities of behavioral circadian rhythms in Drosophila. Cell *99*, 791-802.
46. Shimojo, H., Ohtsuka, T., and Kageyama, R. (2008). Oscillations in notch signaling regulate maintenance of neural progenitors. Neuron *58*, 52-64.

47. Plikus, M.V., and Chuong, C.M. (2008). Complex hair cycle domain patterns and regenerative hair waves in living rodents. J Invest Dermatol *128*, 1071-1080.
48. Westerfield, M. (2000). The zebrafish book. A guide for the laboratory use of zebrafish (Danio rerio). 4th Edition, (University of Oregon press).
49. Holley, S.A., Geisler, R., and Nusslein-Volhard, C. (2000). Control of her1 expression during zebrafish somitogenesis by a delta-dependent oscillator and an independent wave-front activity. Genes Dev *14*, 1678-1690.
50. Julich, D., Hwee Lim, C., Round, J., Nicolaije, C., Schroeder, J., Davies, A., Geisler, R., Lewis, J., Jiang, Y.J., and Holley, S.A. (2005). beamter/deltaC and the role of Notch ligands in the zebrafish somite segmentation, hindbrain neurogenesis and hypochord differentiation. Dev Biol *286*, 391-404.
51. Holley, S.A., Julich, D., Rauch, G.J., Geisler, R., and Nusslein-Volhard, C. (2002). her1 and the notch pathway function within the oscillator mechanism that regulates zebrafish somitogenesis. Development *129*, 1175-1183.
52. Brand, M., Heisenberg, C.P., Jiang, Y.J., Beuchle, D., Lun, K., Furutani-Seiki, M., Granato, M., Haffter, P., Hammerschmidt, M., Kane, D.A., et al. (1996). Mutations in zebrafish genes affecting the formation of the boundary between midbrain and hindbrain. Development *123*, 179-190.
53. Heisenberg, C.P., Brand, M., Jiang, Y.J., Warga, R.M., Beuchle, D., van Eeden, F.J., Furutani-Seiki, M., Granato, M., Haffter, P., Hammerschmidt, M., et al. (1996). Genes involved in forebrain development in the zebrafish, Danio rerio. Development *123*, 191-203.
54. Grandel, H., Lun, K., Rauch, G.J., Rhinn, M., Piotrowski, T., Houart, C., Sordino, P., Kuchler, A.M., Schulte-Merker, S., Geisler, R., et al. (2002). Retinoic acid signalling in the zebrafish embryo is necessary during pre-segmentation stages to pattern the anterior-posterior axis of the CNS and to induce a pectoral fin bud. Development *129*, 2851-2865.
55. Oates, A.C., Mueller, C., and Ho, R.K. (2005). Cooperative function of deltaC and her7 in anterior segment formation. Dev Biol *280*, 133-149.
56. Oates, A.C., and Ho, R.K. (2002). Hairy/E(spl)-related (Her) genes are central components of the segmentation oscillator and display redundancy with the Delta/Notch signaling pathway in the formation of anterior segmental boundaries in the zebrafish. Development *129*, 2929-2946.
57. Zhang, C., Li, Q., Lim, C.H., Qiu, X., and Jiang, Y.J. (2007). The characterization of zebrafish antimorphic mib alleles reveals that Mib and Mind bomb-2 (Mib2) function redundantly. Dev Biol *305*, 14-27.

# Figures

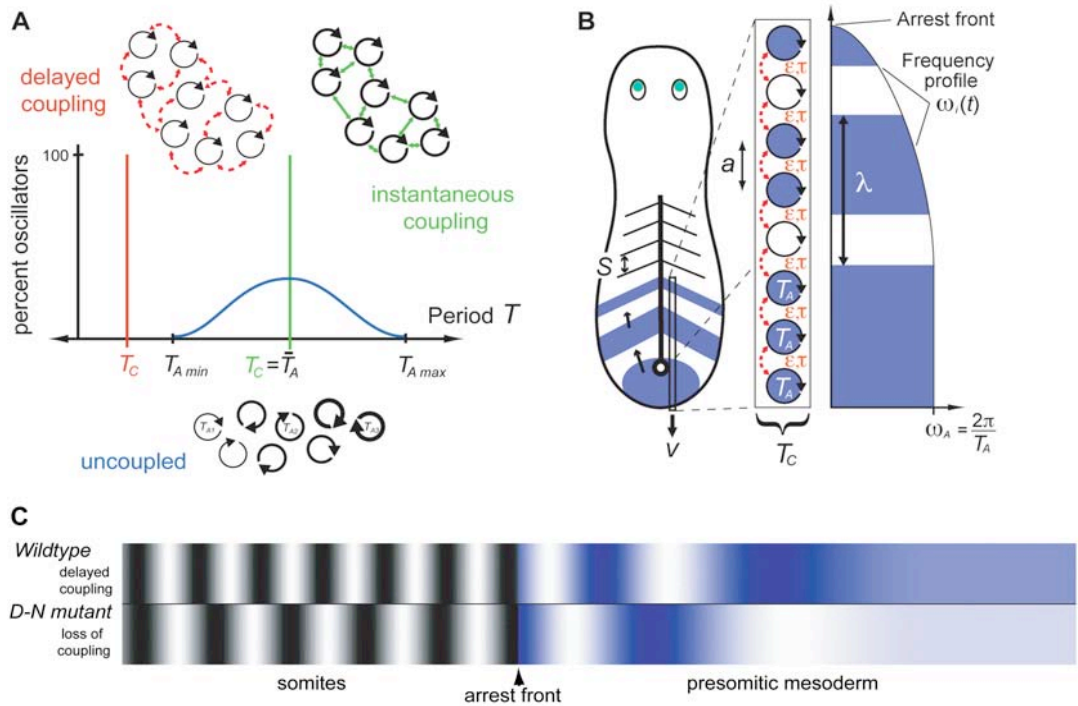

**Figure 1. Theoretical description of collective and spatial properties of oscillators in the segmentation clock**

(**A**). Delayed coupling can alter collective period. Uncoupled oscillators with random phase and unimodal symmetric distribution of autonomous periods $T_{A1}$, $T_{A2}$… (blue line) with average period $\bar{T}_A$. Synchronization by instantaneous weak coupling (green arrows) results in a collective period $T_C = \bar{T}_A$. Synchronization by delayed coupling (red arrows) can result in $T_C$ different to $\bar{T}_A$. Shortening or lengthening of $T_C$ relative to $\bar{T}_A$ is possible depending on the value of the delay. (**B**) The delayed coupling theory describes the segmentation clock as an array of coupled phase oscillators [20]. The key features are: (i) A posterior front describing embryonic elongation with velocity $v$, co-moving with a front that arrests oscillations on the anterior side of the PSM; (ii) Local coupling of oscillators, with strength $\varepsilon$, accounting for Delta-Notch intercellular coupling; (iii) A time delay $\tau$ in coupling, due to synthesis and trafficking of molecules; (iv) A frequency profile $\omega_i(t)$ across the PSM accounting for the slowing of cellular oscillators as they approach the arrest front, characterized by decay-length $\rho$ and the period of the fastest autonomous oscillators $T_A$, located in the posterior PSM. (**C**) Snapshot from Movie S1 generated using the analytical solution to the delayed coupling theory continuum approximation with experimentally determined parameters from this work, predicting that loss of coupling results in increased segment length and wavelength of the oscillatory pattern.

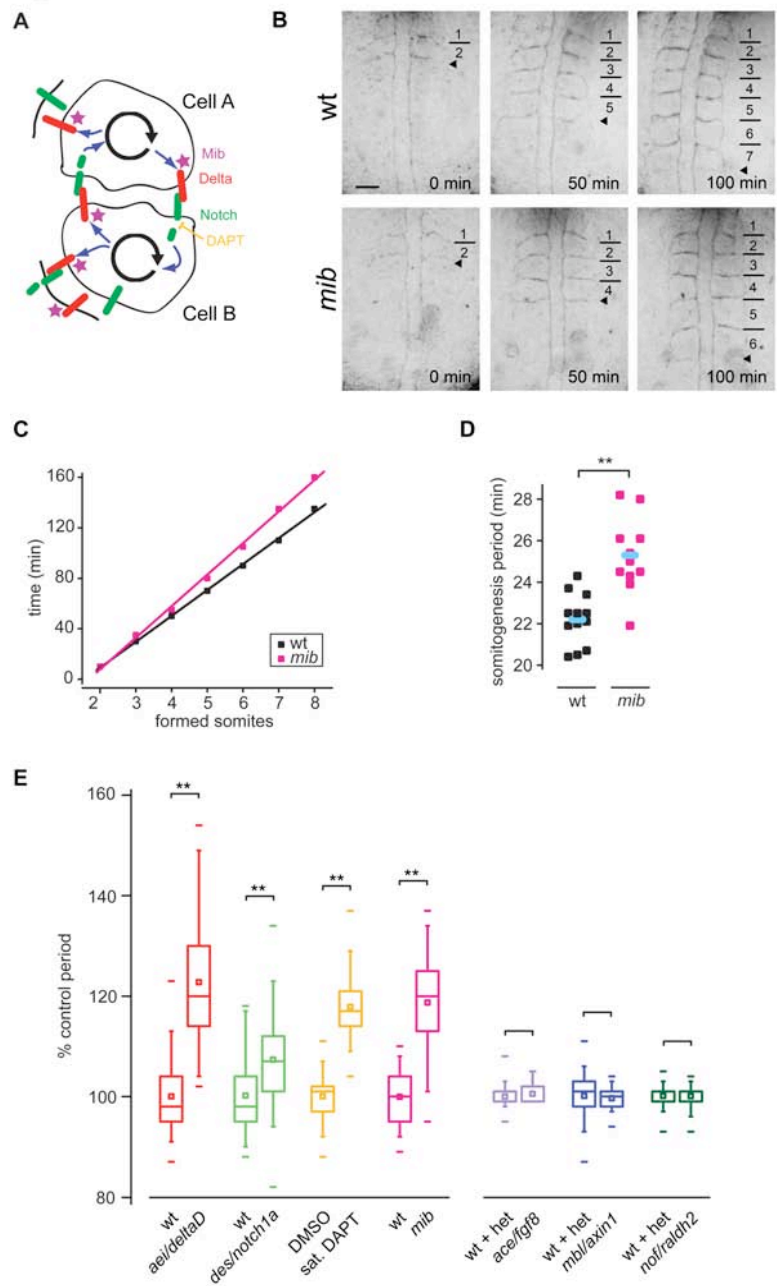

**Figure 2. Somitogenesis period increases after reduction in Delta-Notch coupling**

(**A**) Delta-Notch coupling between two oscillating PSM cells. Delta is the ligand for the Notch receptor, which can be inhibited using the small molecule DAPT. Mib is a ubiquitin ligase required for Delta trafficking and activation. (**B**) Time-lapse movies of wildtype (wt) and *mib* embryos. Bars: formed somite boundaries, arrowheads: forming boundaries. Dorsal view, anterior to top. Scale bar = 50μm. (**C**) Time vs. somite-number plot for (C). Linear fits of data ($R^2$ (wt and *mib*) = 0.999) yield somitogenesis periods of 20.5 min (wt) and 25.0 min (*mib*). (**D**) Distribution of somitogenesis periods in experiment from which (C), (D) were taken, (n(wt) = 12, n(*mib*) = 11). Blue bars: mean somitogenesis period. Temp. = 28.2 ± 0.1°C. (**E**) Box-and-whisker plots of somitogenesis period: n ≥ 37 total embryos, more than six independent trials per experimental condition, except *ace/fgf8*, *mbl/axin1* and *nof/raldh2*: n ≥ 16 total embryos, two independent trials per experimental condition; het, heterozygote. Two asterisks: $p < 0.001$, Student's t-test. Fig. S1 and S2 show that general developmental rate is unaffected in the conditions with slower period.

# Figure 3

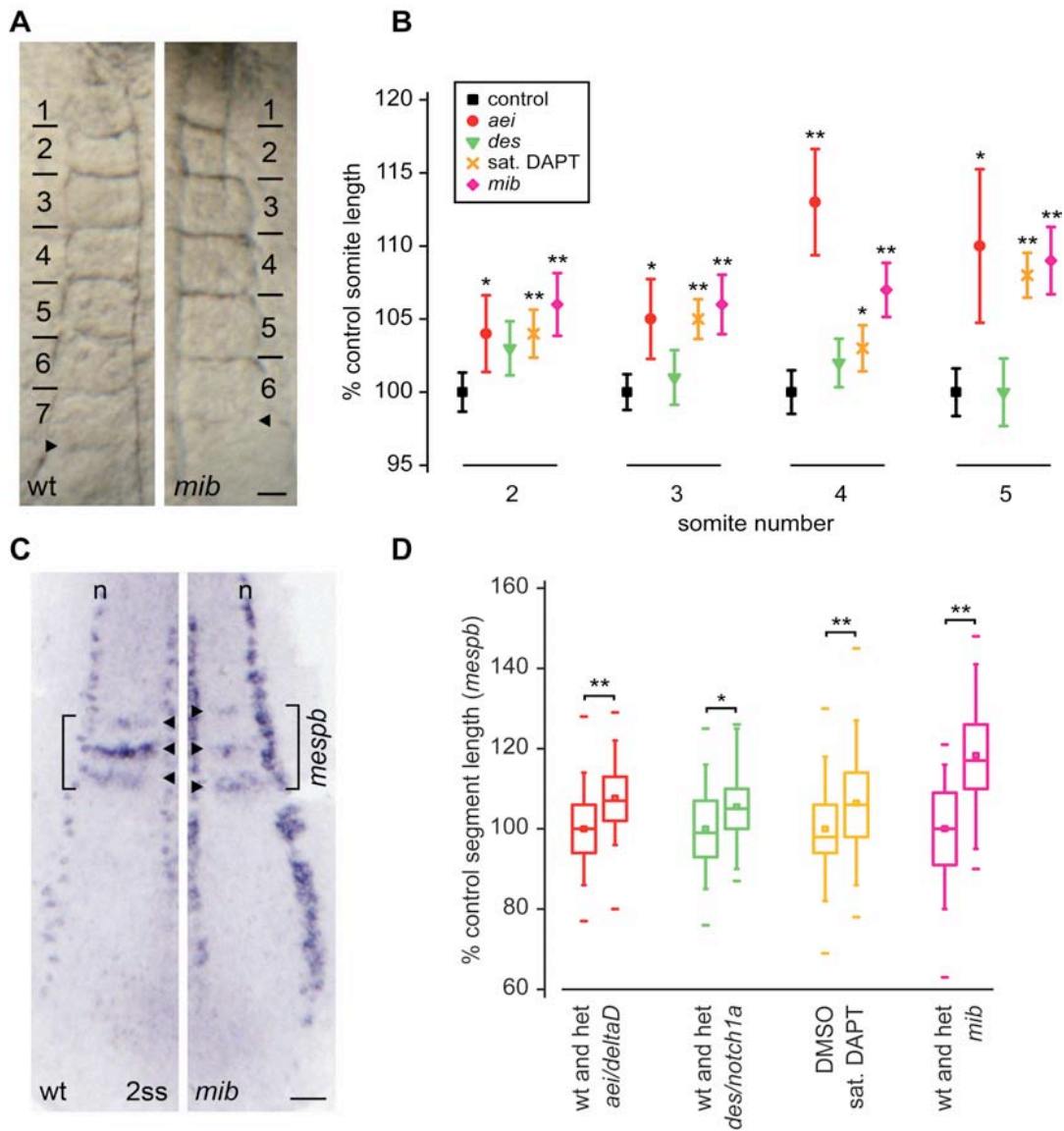

**Figure 3. Segment length increases after reduction in Delta-Notch coupling**

(**A**) Somites of 6-somite-stage live embryos. Bars: formed somite boundaries, arrowheads: forming boundaries. Dorsal view, anterior to top. Scale bar = 25 μm. (**B**) Somite lengths (mean ± 95% CI), n ≥ 40 total embryos per experimental condition, except n = 14 for somite five in *aei/deltaD*, six independent trials per experimental condition. For control population, largest CI detected is displayed. (**C**) *In situ* hybridization *mespb* (arrowheads), *isl1* (interneurons, Rohon-Beard neurons, n). Scale bar = 50 μm. (**D**) Box-and-whisker plots of segment length, n ≥ 40 total embryos, more than three independent trials per experimental condition. One and two asterisks: $p < 0.01$, $p < 0.001$, Student's t-test. Fig. S3 shows that the position of the arrest front in the PSM is unchanged in the conditions with increased segment length.

Figure 4

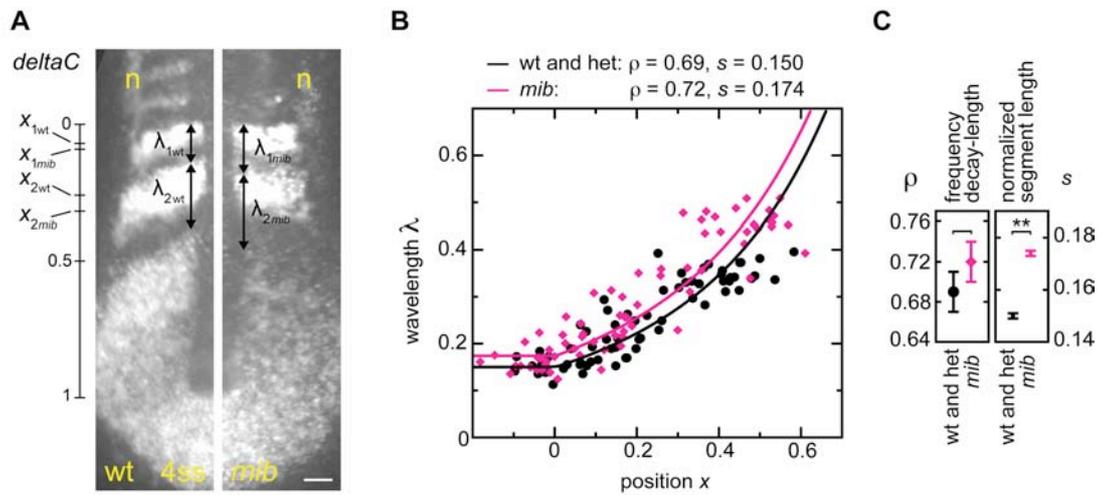

**Figure 4. Determination of segmentation clock collective period from oscillating gene expression patterns**

(**A**) *In situ* hybridization: *deltaC* and *isl1* (interneurons and Rohon-Beard neurons, n) in representative wildtype and *mib* embryos. Scale bar = 50 μm. (**B**) Measurements of normalized gene expression wavelength $\lambda$ and position $x$; wildtype black, *mib* pink. Data points: n(wt) = 65, 28 embryos, n(*mib*) = 70, 28 embryos, two independent trials. Curves show fit of Eq. (3) to data (Experimental Procedures). (**C**) Values of frequency profile decay-length $\rho$, and normalized segment length $s$ parameters from the fit to the data. Error bars: 95% CI from bootstrap analysis (Supplementary Experimental Procedures 2.2.5). Two asterisks: $p < 0.001$, Student's t-test.

# Figure 5

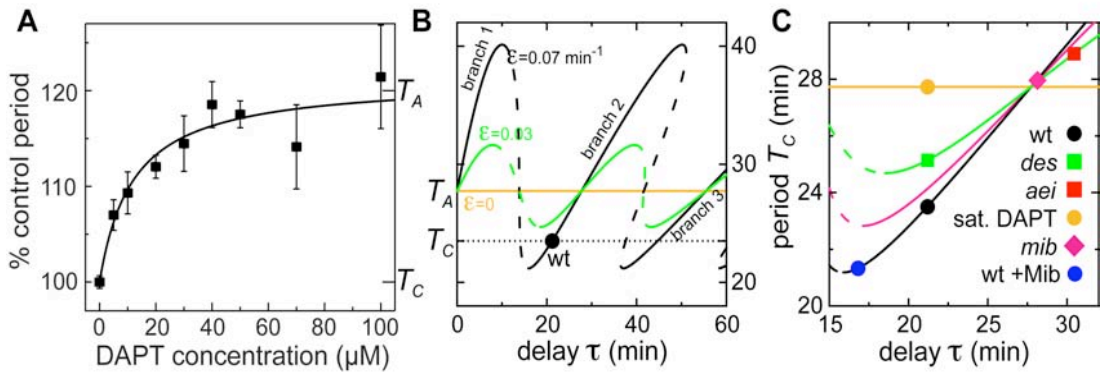

**Figure 5. Coupling strength and time delays regulate the collective period of the segmentation clock**

(**A**) Somitogenesis period (mean±95%CI) vs. DAPT concentration: n≥38, more than four independent trials for data points except 70 and 100μM: n≥18, two independent trials. Curve: fit to the delayed coupling theory yielding $\varepsilon = 0.07 \pm 0.04$ and $\tau = 21 \pm 2$ min (Eq. (4), Experimental Procedures). (**B**) Diagram of solutions to Eq. (1). Lines: collective period $T_C$ vs. coupling delay $\tau$ for different values of coupling strength $\varepsilon$, as indicated. Stable and unstable solutions: solid and dashed lines, respectively. Dotted line: wildtype $T_C$, 28°C. (**C**) Close-up of (B): approximate positions of experimental conditions in parameter space (Supplementary Experimental Procedures 2.2.7). Black dot: wildtype, 10 somite stage, 28°C; blue dot refers to experiments with elevated Mib levels in Fig. 6. Because the Anterior Limit of Defects (ALD) of *mib* is between those of wildtype and *des/notch1a*, its coupling strength $\varepsilon$ is assumed to be between wildtype and *des/notch1a*. The ALD of *aei/deltaD* is similar to *des/notch1a*, hence their coupling strengths are similar. Lines: $T_C$ vs. $\tau$ relation for: black $\varepsilon=0.07$ min$^{-1}$, pink $\varepsilon=0.05$ min$^{-1}$, green $\varepsilon=0.03$ min$^{-1}$, orange $\varepsilon=0$ min$^{-1}$. Fig. S4. shows that *mib* heterozygote embryos have a longer period.

Figure 6

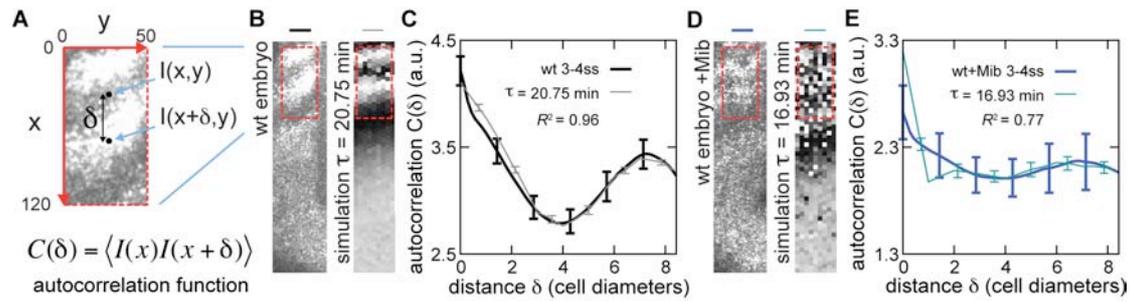

**Figure 6. Time delays regulate the stability of the segmentation clock**

(**A**) Schematic of how the autocorrelation function (Eq. (6), Experimental Procedures) is computed. A magnification of the gene expression pattern from B is shown, together with the reference axes *x* and *y* in pixels, and the distance *d* between two sample points of gene expression intensity *I*. (**B**) Representative cyclic *dlc* expression in PSM of wildtype embryo and DCT simulation with wildtype parameters defined in this work, black dot in Fig. 5C. (**C**) Average autocorrelation function of spatial patterns in red box from B: n(embryos) = 15, black line; n(simulations) = 20, grey line. (**D**) Representative experimental Mib over-expression (400 pg mRNA) and corresponding DCT simulation with reduced coupling delay, blue dot in Fig. 5C. (**E**) Average autocorrelation function of spatial patterns in red box from D: n(embryos) = 8, dark blue line; n(simulation) = 20, light blue line. The arbitrary units in the correlation axis are multiplied by $10^{-3}$. Fig. S5 and S6 show the autocorrelations for decreasing delays in numerical simulations and increasing levels of Mib over-expression in the embryo, and Fig. S7 shows that the Delta-Notch loss of coupling mutants have distinct autocorrelation functions.

*Tables*

**Table 1. Segmentation variables in Delta-Notch mutant and DAPT-treated embryos**

| Experimental condition | % control period [b] (timelapse analysis) | % control somite length [c] (5th somite) | % control segment length [d] (*mespb* pattern) | % control period [e] (*deltaC* stripe wavelength) | Anterior Limit of Segmental Defects |
|---|---|---|---|---|---|
| *aei/deltaD* | 123 ± 4 [a] | 110 ± 5 | 108 ± 2 | n.d [f]. | 7 ± 2 [31] |
| *des/notch1a* | 107 ± 3 | 100 ± 2 | 105 ± 3 | n.d. | 7 ± 2 [13, 31] |
| sat. DAPT | 118 ± 1 | 108 ± 2 | 106 ± 2 | n.d. | 5.2 ± 0.2 [13] |
| *mib* | 119 ± 2 | 109 ± 2 | 118 ± 5 | 116 ± 1 | 10 – 12 [57] |

[a] Mean ± 95% CI. [b] Period measurements: *aei/deltaD*: n(wt) = 46, n(mutant) = 50, 6 independent trials; *des/notch1a*: n(wt) = 37, n(mutant) = 49, 6 independent trials; saturating DAPT: n(DMSO) = 114, n(sat. DAPT) = 141, 12 independent trials; *mib*: n(wt) = 64, n(mutant) 73, 8 independent trials. [c] Somite length measurements (5th somite): *aei/deltaD*: n(wt and het) = 78, n(mutant) = 14; *des/notch1a*: n(wt and het) = 112, n(mutant) = 43; sat. DAPT: n(DMSO) = 104, n(sat. DAPT) = 74; *mib*: n(wt and het) = 91, n(mutant) = 47, 6 independent trials for all experimental conditions. [d] Segment length measurements: *aei/deltaD*: n(wt and het) = 132, n(mutant) = 69, 4 independent trials; *des/notch1a*: n(wt and het) = 80, n(mutant) = 41, 3 independent trials; sat. DAPT: n(DMSO) = 155, n(sat. DAPT) = 239, 8 independent trials; *mib*: n(wt and het) = 40, n(mutant) = 40, 4 independent trials. [e] Stripe wavelength measurements: *mib*: n(wt and het) = 65 data points from 28 embryos, n(mutant) = 70 data points from 28 embryos, 2 independent trials. The largest confidence intervals in control populations were 100 ± 2 for period and somite length, 100 ± 4 for segment length, and 100 ± 4 for stripe wavelength in *mib*. [f] n.d., not done.

Supplemental Information for

# Intercellular coupling regulates the period of the segmentation clock

Leah Herrgen[1], Saúl Ares[2], Luis G. Morelli[1,2,3], Christian Schröter[1], Frank Jülicher[2], Andrew C. Oates[1]*

[1] Max Planck Institute for Molecular Cell Biology and Genetics, Pfotenhauerstr 108, 01307 Dresden, Germany.
[2] Max Planck Institute for the Physics of Complex Systems, Nöthnitzer Straße 38, 01187 Dresden, Germany.
[3] Departamento de Fisica, FCEyN, UBA, Pabellón I, Ciudad Universitaria, Buenos Aires, Argentina.
* Author for correspondence: oates@mpi-cbg.de

# Contents

## *1. Supplemental Data*

## *2. Supplemental Experimental Procedures*

## *3. Supplemental References*

# 1. Supplemental Data

## 1.1 Supplemental Data relating to main text Figure 2.

### 1.1.1 General developmental progression is unchanged in Delta-Notch mutant and DAPT-treated embryos

In order to assess whether general developmental progression and tissue differentiation were affected in our experimental conditions, we analyzed expression patterns of *egr2b* (*krox20*) which marks the 3[rd] and 5[th] rhombomere [1], and of the markers of paraxial mesoderm *spt/tbx16* [2] and intermediate mesoderm *gata1* [3]. Expression of all markers was unchanged in *mib* (Fig. S1), *aei/deltaD and des/notch1a* embryos, and in embryos treated with saturating concentrations of DAPT, as compared to their wildtype/heterozygous or DMSO-treated control siblings at the 10 somite stage (data not shown). We conclude that general developmental progression and tissue differentiation are not detectably affected in the relevant experimental conditions, in agreement with previous studies [4, 5].

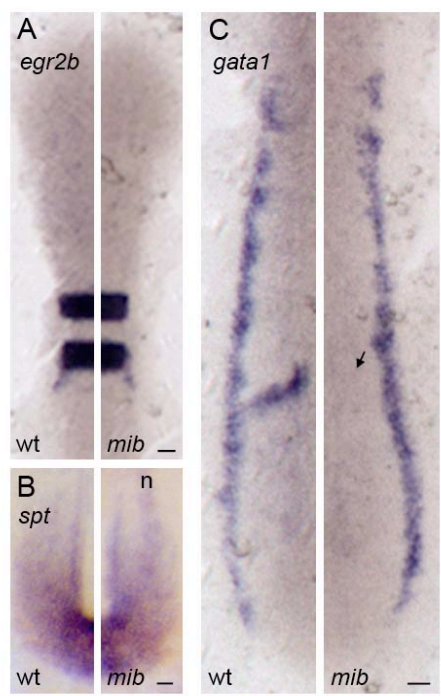

**Figure S1. Expression of mesoderm and ectoderm differentiation markers in *mib* embryos**
*In situ* hybridization with ectoderm **(A)** and mesoderm **(B, C)** markers in the PSM of wildtype/heterozygous and *mib* embryos at the 10 somite stage. Elevated expression of *isl1* in (B) in interneurons and Rohon-Beard neurons (n, out of focal plane) and reduced expression of *mespa* (arrow in C) is observed in *mib* embryos, either of which was used to distinguish wildtype/heterozygous from *mib* embryos. Dorsal view with anterior to the top. Expression patterns were assessed in at least five embryos per experimental condition. Scale bars = 25 μm.

### 1.1.2 *Rate of axial elongation is unchanged in Delta-Notch mutant and DAPT-treated embryos*

Arrest front velocity is one of the key parameters of the delayed coupling theory (DCT), as it controls segment length together with collective period of the clock according to $S = v\, T_C$ [6, 7]. This predicts that changes to clock period will give rise to correspondingly longer segments, if arrest front velocity remains unchanged. Arrest front velocity depends on (i) the rate of arrest front movement with axial elongation and (ii) the rate of arrest front movement within the PSM (see also Supplemental Data 1.1.3).

To determine the rate of axial elongation, we measured embryo length at different stages in time-lapse movies of control and mutant or DAPT-treated embryos [8, 9] (Fig. S2). Embryo length was measured along the yolk from the anterior end of the embryo to the anterior end of the paraxial mesoderm, and along the dorsal side of the paraxial mesoderm until the posterior end of the embryo (Fig. S2A). This procedure proved to be robust with respect to slightly different mounting orientations, and applicable to embryos of all relevant stages. Although the absolute size of embryos in a clutch was variable between different trials depending on the egg size from the mother, we did not detect any difference in embryo length between control and experimental conditions at any time point (Fig. S2B-E and Movies S4 and S5).

We conclude that axial elongation, one of the controlling parameters of arrest front velocity, is not detectably changed in the relevant experimental conditions. Furthermore, these results also indicate that general developmental progression is unimpaired, in accordance with results from gene expression studies (Fig. S1).

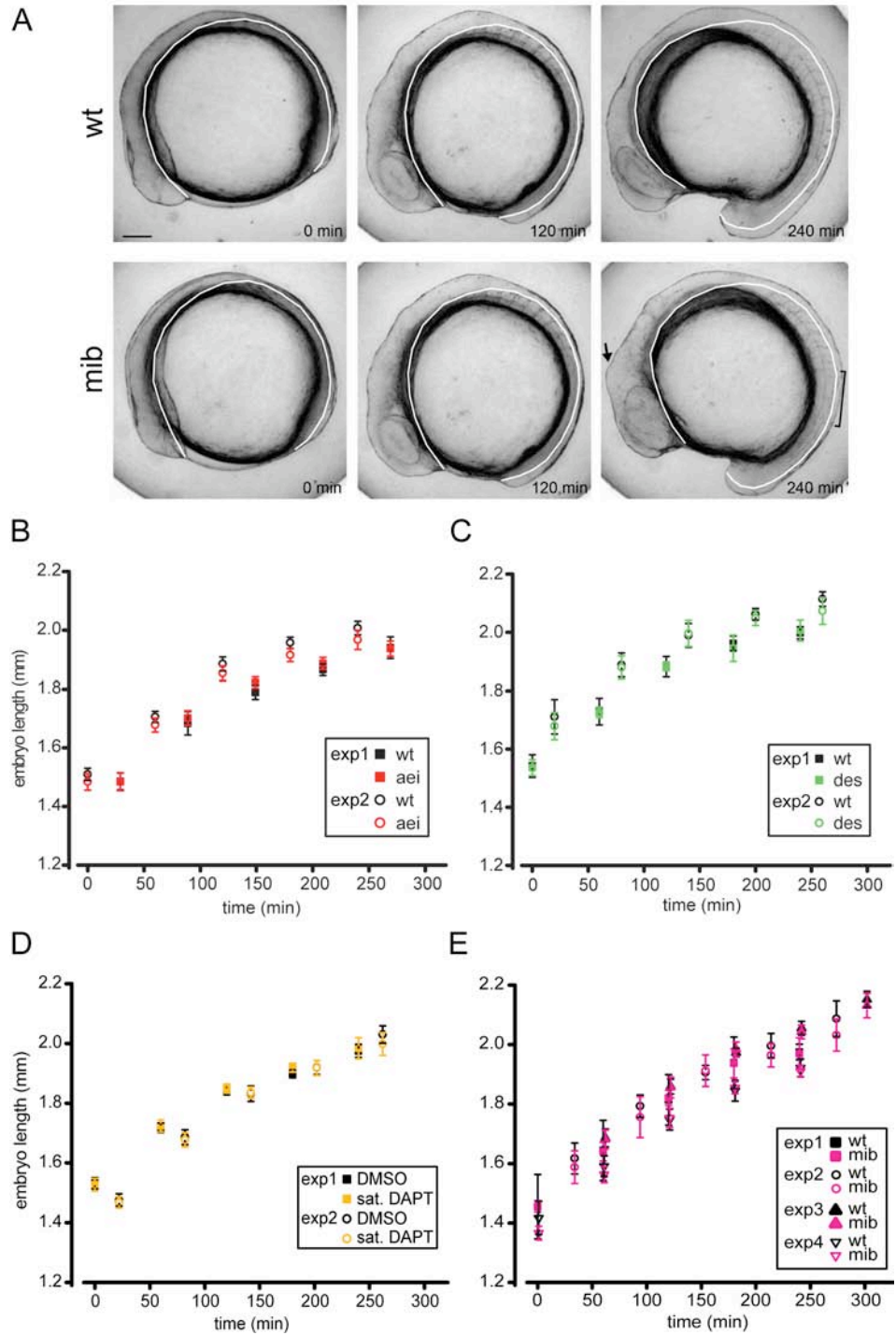

**Figure S2. Axial elongation is not changed in Delta-Notch mutant and DAPT-treated embryos**
**(A)** Stills from time-lapse movies [8, 9] of wildtype and *mib* embryos. White lines: line along which axial length was measured. Arrow: bulge in mutant brain due to neuronal hyperplasia. Bracket: region containing irregular somite boundaries in the mutant. Lateral view with anterior to the left. Scale bar = 100 μm. **(B-E)** Plots of embryo length (mean ± 95% confidence interval, CI) vs. elapsed recording time for different experimental conditions, n ≥ 4 embryos for each data point. The time points at which wildtype embryos reached the 4 or 10 somite stage are indicated. Values for control and mutant or DAPT-treated embryos were not significantly different ($p > 0.01$ in all cases as assessed by Student's t-test) at any time point or for any of the experimental conditions.

## 1.2   Supplemental data relating to main text Figure 3.

### 1.2.1 Position of arrest front is unchanged within PSM of Delta-Notch mutant and DAPT-treated embryos

The position of the arrest front was proposed to correspond to the position of the posterior boundary of *Mesp2* expression in the mouse PSM [10]. Accordingly, we identify the position of the arrest front with the position of the posterior boundary of the *mespb* expression domain in zebrafish, and define *d* as its distance to the posterior tip of the embryo (Fig. S3A). The position of the arrest front moves posteriorly within the PSM during the process of somitogenesis, thereby considerably shortening this distance (Fig. S3B-E). Arrest front velocity is therefore determined by the rate of arrest front movement within the PSM as well as by the rate of axial elongation (see Supplemental Data 1.1.2).

We measured *d* in Delta-Notch mutants and DAPT-treated embryos at 3-4 time points spanning the developmental interval from the formation of somite 1 to somite 12, depending on the experimental condition (Table S1). We detected no difference in *d* between control and experimental conditions at any time point (Fig. S3B-E). Combined with the findings on the rates of axial elongation (Supplemental Data 1.1.2), we conclude that arrest front velocity is not detectably changed in the relevant experimental conditions. Thus, these findings rule out potential alternative explanations for the observed concomitant increases in somitogenesis period and segment length that involve significant changes to the arrest front velocity. The conclusion of normal arrest front velocity is also necessary for the independent estimates of an increase in clock collective period from cyclic gene expression stripe wavelength measurements in Fig. 4 in the main text.

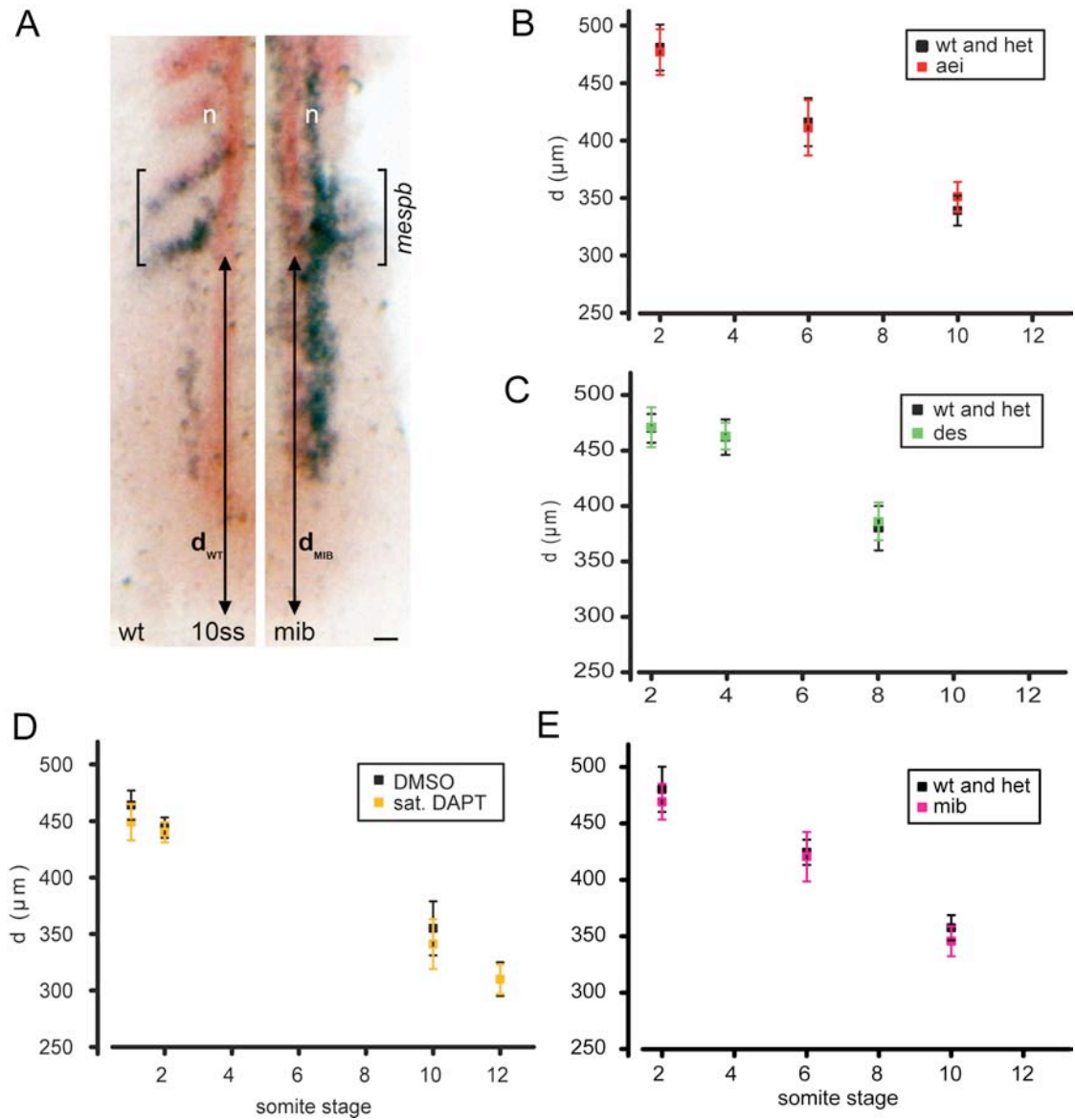

**Figure S3. Position of arrest front is unchanged within the PSM of Delta-Notch mutants and DAPT-treated embryos**

**(A)** *In situ* hybridization for *mespb* (blue) and *isl1* (blue; expressed in interneurons and Rohon-Beard neurons, n) and *myoD* (red) in the PSM of wildtype/heterozygous and *mib* embryos. Bracket: expression pattern of *mespb*, striped in wildtype and in a salt-and-pepper pattern in *mib*. Double-headed arrow: d, distance between the posterior boundary of *mespb* expression domain and the posterior tip of the embryo. Dorsal view with anterior to the top. Scale bar = 25 μm. **(B-E)** Plots of d (mean ± 95% CI) vs. developmental stage in different experimental conditions, $10 \leq n \leq 59$ (mean = 22.5) embryos from at least two independent trials for each data point (Table S1). Each experimental condition had three trials for at least one of the data points. Somite stages are given for the wildtpye siblings or clutch mates. Values of arrest front position for control and mutant or DAPT-treated embryos were not significantly different ($p > 0.01$ in all cases as assessed by Student's t-test) at any time point or for any of the experimental conditions.

**Table S1.** Experiment and embryo numbers for measurement of the position of the determination front in the PSM

| experimental condition | stage (# somites) | independent trials | n(wt) | n(mut) |
|---|---|---|---|---|
| *aei/deltaD* | 2 | 3 | 30 | 18 |
| | 6 | 2 | 20 | 15 |
| | 10 | 2 | 20 | 20 |
| *des/notch1a* | 2 | 3 | 30 | 20 |
| | 4 | 2 | 19 | 20 |
| | 8 | 2 | 20 | 19 |
| *mib* | 2 | 3 | 23 | 23 |
| | 6 | 2 | 20 | 16 |
| | 10 | 2 | 20 | 14 |
| DAPT | 1 | 2 | 20 | 14 |
| | 2 | 3 | 29 | 52 |
| | 6 | 2 | 10 | 16 |
| | 10 | 3 | 29 | 59 |

## 1.3 Supplemental data relating to main text Figure 5

### 1.3.1 Somitogenesis period of heterozygous Delta-Notch mutants

In order to investigate whether heterozygote mutant embryos showed intermediate somitogenesis periods, we identified the heterozygotes by PCR genotyping. We did not detect statistically significant period changes in *aei/deltaD* and *des/notch1a* heterozygous embryos, but found a small increase in *mib* heterozygotes (Fig. S4). The observed change was almost an order of magnitude smaller than that observed in homozygous *mib* mutants, and we therefore decided to include heterozygous embryos in control populations of segment length experiments. Although variability of the control population must consequentially increase, inclusion of heterozygotes in the control population considerably simplified experimental procedures.

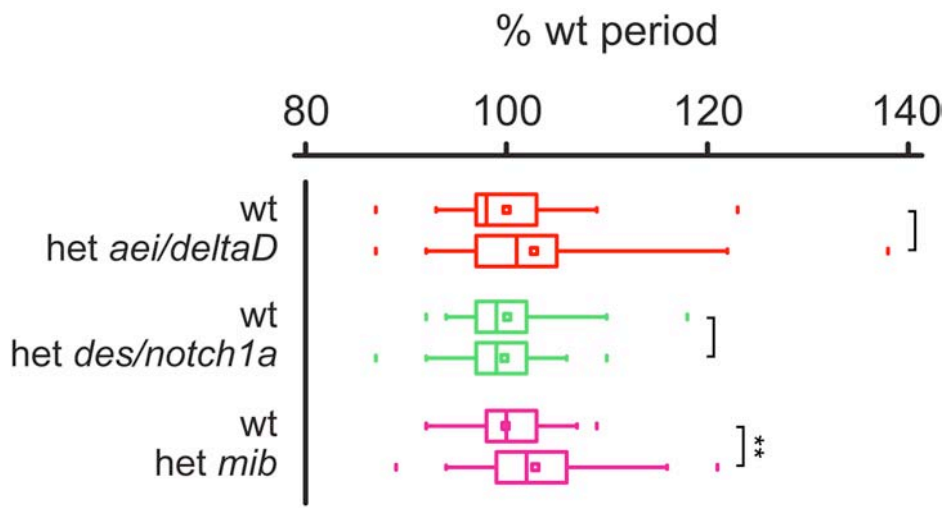

**Figure S4. Somitogenesis period of heterozygous Delta-Notch mutants**

Box plots of somitogenesis period determined from multiple-embryo time-lapse movies [8, 9] for populations of heterozygous Delta-Notch mutants and their wildtype sibling embryos. n ≥ 37 total embryos from at least six independent trials per experimental condition. Box plots: The central box covers the interquartile range with the mean indicated by the small square and the median by the line within the box. The whiskers extend to the $5^{th}$ and $95^{th}$ percentiles, and small bars depict the most extreme values. Two asterisks: $p < 0.001$ as assessed by Student's t-test.

### *1.4 Supplemental data relating to main text Figure 6*

#### *1.4.1 Autocorrelations are lost as the coupling delay is decreased in numerical simulations of the PSM*

The delayed coupling theory (DCT) predicts the existence of instabilities for certain ranges of the coupling delay, depending also on coupling strength [7]. From the parameters that we have estimated in this work we predict that the segmentation clock of the wildtype zebrafish operates not far from such an instability, see Fig. 5B,C in main text. We first explored the effects of this instability in the delayed coupling theory using numerical simulations of segmentation on a two dimensional hexagonal lattice of cells. The simulations were carried out as described in Supplemental Experimental Procedures section 2.2.9 with parameter values determined in this work for the simulation of wildtype: 1 cell diameter = 7 pixels, $\sigma$ = 27 cell diameters, $\omega_A$ = 0.2205 $min^{-1}$, $\varepsilon$ = 0.07 $min^{-1}$, $\tau$ = 20.75 min, $v$ = 0.249 cell diameters/min (Experimental Procedures, and Supplemental Experimental Procedures 2.2.4, and 2.2.6). Fluctuations were introduced with a noise term with strength $\eta$ = 0.02 $min^{-1}$. From these simulations we obtained an average autocorrelation function describing noisy gene expression patterns (Supplemental Experimental Procedures 2.2.10) in excellent agreement with that obtained from the oscillatory expression pattern of *deltaC* in wildtype embryos (Fig. S5A; main text Fig. 6B,C). The wildtype autocorrelations show a valley at 4 cell diameters, which marks the typical distance of an interstripe, and then a peak at ~7 cell diameters, which gives the stripe wavelength. Because these autocorrelations are recorded in the anterior PSM, this distance is very close to the segment length.

We next decreased the value of the time delay in the simulations without changing the rest of the parameters. As the coupling delay is decreased, an increasingly noisy pattern is observed (Fig. S5A-E). We characterized the noise in the patterns by means of the autocorrelation function without scaling or offset. We found that the average autocorrelation function progressively loses the structure characterizing the inter-stripe and stripe wavelength, reflecting an increasing level of disorder in the phase pattern and the emergence of correlations at intermediate length-scales (Fig. S5F-J). Due to the complex interplay of noise and delayed coupling close to the stability boundary, the signatures of the instability are observed before the stability boundary is actually crossed at delays of about 16 min. In the instability, short range anti-correlations are observed due to partial frustration among neighboring oscillators. This is in qualitative agreement with spatial patterns observed in simulations of genetic network models of the segmentation clock for some values of the coupling delay [11, 12].

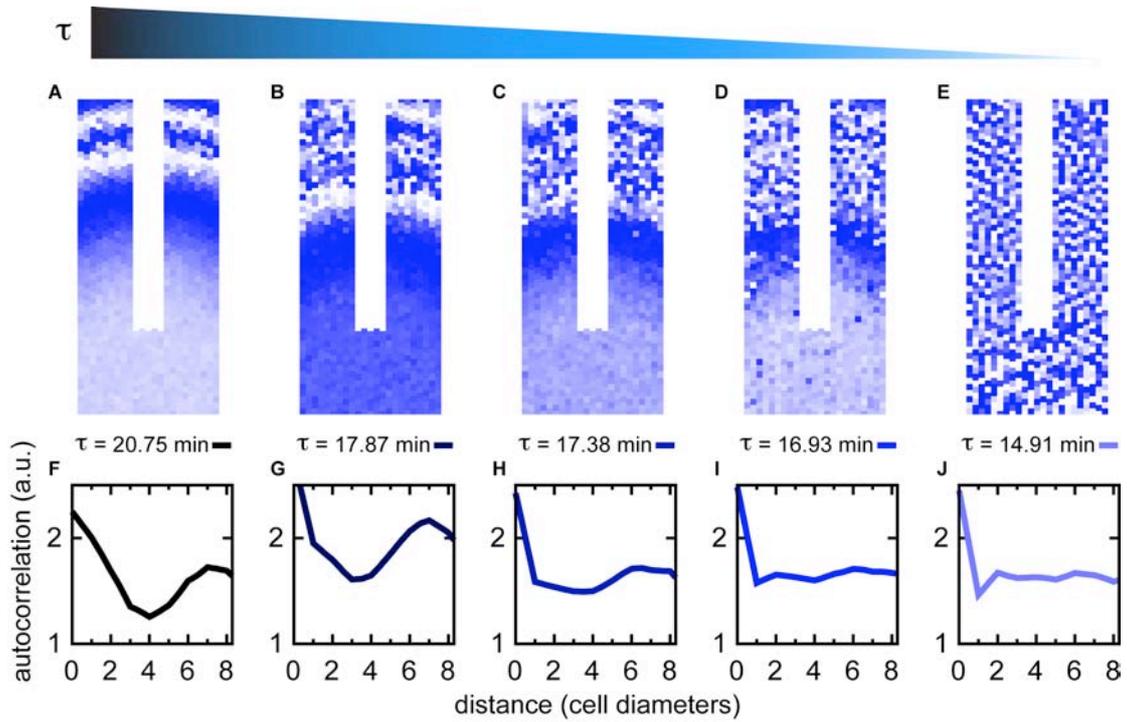

**Figure S5. Autocorrelations are progressively lost as the coupling delay is decreased in numerical simulations of the PSM.**

**(A-E)** Snapshots of numerical simulations and **(F-J)** plots of the average autocorrelation function for numerical simulations of the theory with **(A, F)** wildtype parameters and **(B-E, G-J)** decreasing coupling delay. The autocorrelation was first computed for individual simulations in which the intensity of the color maps the sine of the phase with values from 1 to -1 [7], and then averaged as described in Supplemental Experimental Procedures 2.2.10. The arbitrary units in the correlation axis are multiplied by $10^{-4}$. As coupling delay $\tau$ is reduced from wildtype values $\tau \sim 21$ min, the system approaches the stability boundary at $\tau \sim 16$ min. Progressively disrupted patterns are observed, and structures are formed with length-scales that span from one to several cell diameters. This leads to a loss of the one segment-length correlations observed for wildtype parameters. Number of single realizations is 20 for each plot. The arbitrary units in the correlation axis are multiplied by $10^{-3}$.

### 1.4.2 Increasing **Mib** *levels produce increasingly disrupted patterns of cyclic gene expression consistent with reduced coupling delay*

To test the delayed coupling theory's prediction about the existence of an instability (Supplemental Data 1.4.1), we injected *mib* mRNA into wildtype embryos at the one-cell stage and examined the cyclic gene expression patterns of *deltaC* at the 3-4 somite stage. We used the 3-4 somite stage because the mRNA injection phenotype was strongest and maximally penetrant at this time. The effects were usually restricted to the left- or right-hand side of the embryo due to unequal distribution of mRNA, as expected for zebrafish microinjection and followed by tracer GFP mRNA, and we analyzed the side with the stronger defects. We found that with increasing levels of *mib* mRNA, the gene expression patterns appear more frequently and increasingly disrupted, with characteristic defects that span a range of length-scales as reflected in the gradual flattening of the average autocorrelation functions shown in Figure S6A-F. This demonstrates the value of a quantitative tool to assess the gradual effects of increasing concentrations of a treatment from very subtle to strong effects. In addition, a direct comparison with the result of numerical simulations (Fig. S5) is possible. The average autocorrelation functions show that the effects of increasing Mib levels are consistent with a decreasing coupling delay (directly compared in main text Fig. 6E for 400 pg *mib* mRNA). The upper limit of mRNA injection was 500 pg, since at this level, abnormal appearing embryos were occasionally observed and had to be excluded from analysis. Yet higher levels of *mib* mRNA resulted in reduced embryo viability and were not analyzed. Control experiments were carried out injecting the same concentration range of GFP mRNA. The resulting autocorrelation functions are shown in Figure S6G-K. GFP injected embryos show average autocorrelation functions similar to wildtype, and thus rule out the possibility that the effect observed in the Mib injected embryos is a generic artifact of the injection protocol. Although a decrease of coupling delays is also predicted to decrease the collective period, in practice increasingly elevated levels of Mib resulted in increasingly abnormal somitogenesis, preventing the measurement of period.

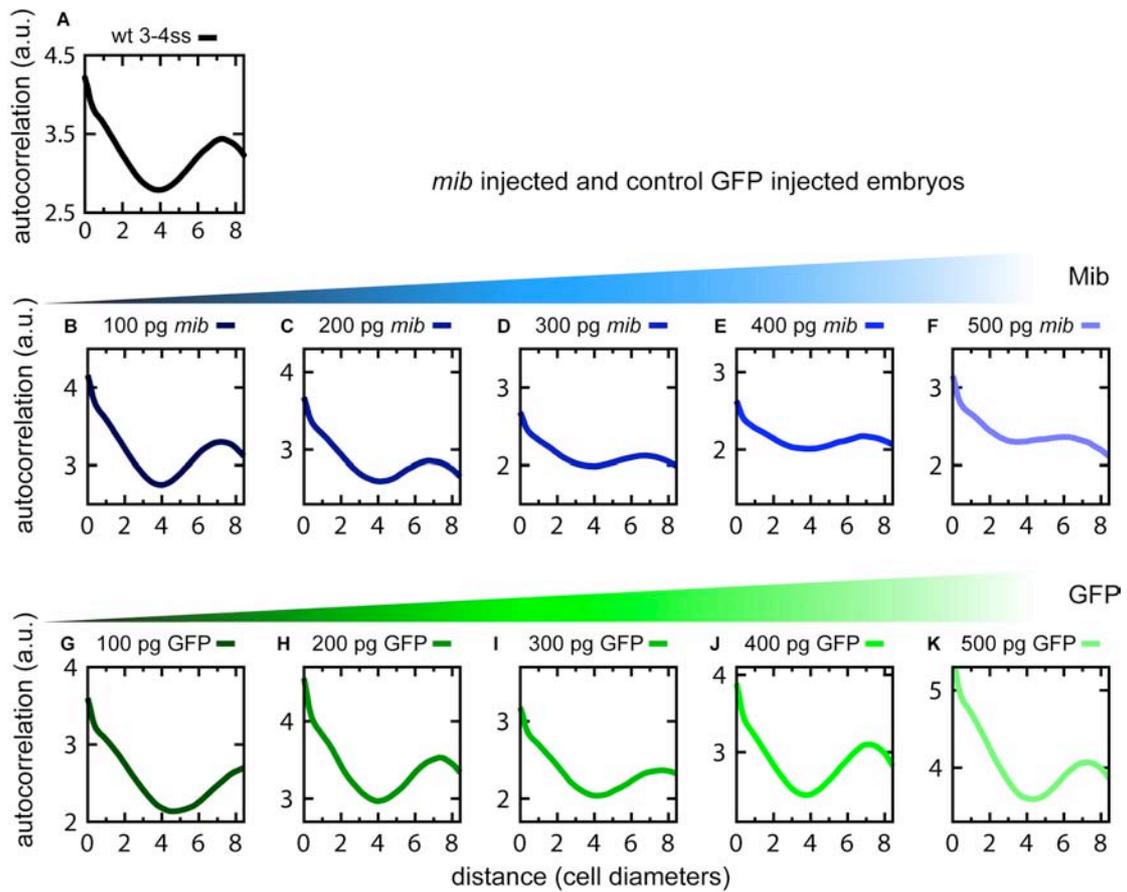

**Figure S6.** *Increasing* **Mib** *levels produce increasingly disrupted patterns of cyclic gene expression consistent with reduced coupling delay.*

Average autocorrelations of *deltaC* gene expression patterns in wildtype embryos fixed at 3-4 ss (**A**), *mib* mRNA over-expressing embryos **(B-F),** and control experiments injecting GFP mRNA at the same concentrations **(G-K)**. Average autocorrelation functions of Mib-expressing embryos reflect the fact that gene expression patterns are increasingly disrupted as Mib levels are increased. The autocorrelations are consistent with those of decreasing coupling delays, see Fig. S5F-J and Fig. 6D,E of the main text. GFP mRNA injected embryos display average autocorrelation functions similar to wildtype. **(A)** wildtype embryo. **(B)** 100 pg *mib* (n=14), **(C)** 200 pg *mib* (n=17), **(D)** 300 pg *mib* (n=11), **(E)** 400 pg *mib* (n=8), and **(F)** 500 pg *mib* (n=9). **(G)** 100 pg GFP (n=10), **(H)** 200 pg GFP (n=10), **(I)** 300 pg GFP (n=10), **(J)** 400 pg GFP (n=10), and **(K)** 500 pg GFP (n=9). The arbitrary units in the correlation axis are multiplied by $10^{-3}$.

### 1.4.3 Cyclic gene expression patterns resulting from elevated Mib levels are distinct from loss of coupling mutants

Disruptions to the wildtype *deltaC* oscillating gene expression pattern resulting from elevated Mib levels are distinctive and differ from other conditions that induce failure of the clock, for example the loss of coupling mutants. While the typical gene expression pattern of these mutants is very noisy, with a fine structure of subcellular length scale and a local uncorrelated salt-and-pepper pattern, the patterns we observed after Mib overexpression show larger structures ranging across several cell diameters. We quantitatively examined the oscillating gene expression patterns in loss of coupling mutants *aei/deltaD*, *des/notch1a*, and *mib* using the autocorrelation function, and compared these results to the wildtype and Mib over-expressing embryos (Fig. S6). For the comparison between wildtype and Mib-overexpression, we used the 3-4 somite stage because the mRNA injection phenotype was maximally penetrant at this time. For the loss of coupling mutants, we focused on later stages of development where the desynchronized state is reached.

The autocorrelation function clearly shows that the loss of coupling mutants are similar to each other, but different to the Mib over-expression phenotype and to the wildtype. We ignore correlations for length scales less than one cell, which reflect the heterogeneous sub-cellular distribution of *deltaC* mRNA. As mentioned in section 1.4.1, the wildtype correlations show a valley at approximately 4 cell diameters, which marks the typical distance of an interstripe and a peak at just above 7 cell diameters, which gives the stripe wavelength (Fig. S7A, C, E, I). Because these correlations are recorded in the anterior PSM, this distance is very close to the segment length. In contrast, loss of coupling mutants have a flat correlation peak at ~ 3 cell diameters and a continuous decrease in correlations at longer distances, which reflects the expected salt-and-pepper scatter of desynchronized oscillating cells and the amplitude modulation between the anterior and posterior domains of the PSM (Fig. S7F-H, J-L). The pattern of the Mib over-expressing embryos is different again (Fig. S7B, D): here the increased correlation at the distance where the wildtype has the interstripe valley reflects the existence of gene expression structures at these intermediate sizes, and the peak at just below 7 cell diameters reveals the existence of a spatial wavelength similar to wildtype (Fig. S7D). Combined, these results indicate that the over-expression of Mib results in perturbations to the oscillating gene expression patterns that are quantitatively distinguishable from wildtype and loss of coupling situations. Furthermore, they show that the autocorrelation function is a simple and powerful method to distinguish different experimental conditions that perturb the segmentation clock, and also to compare the experimental conditions with theoretical predictions.

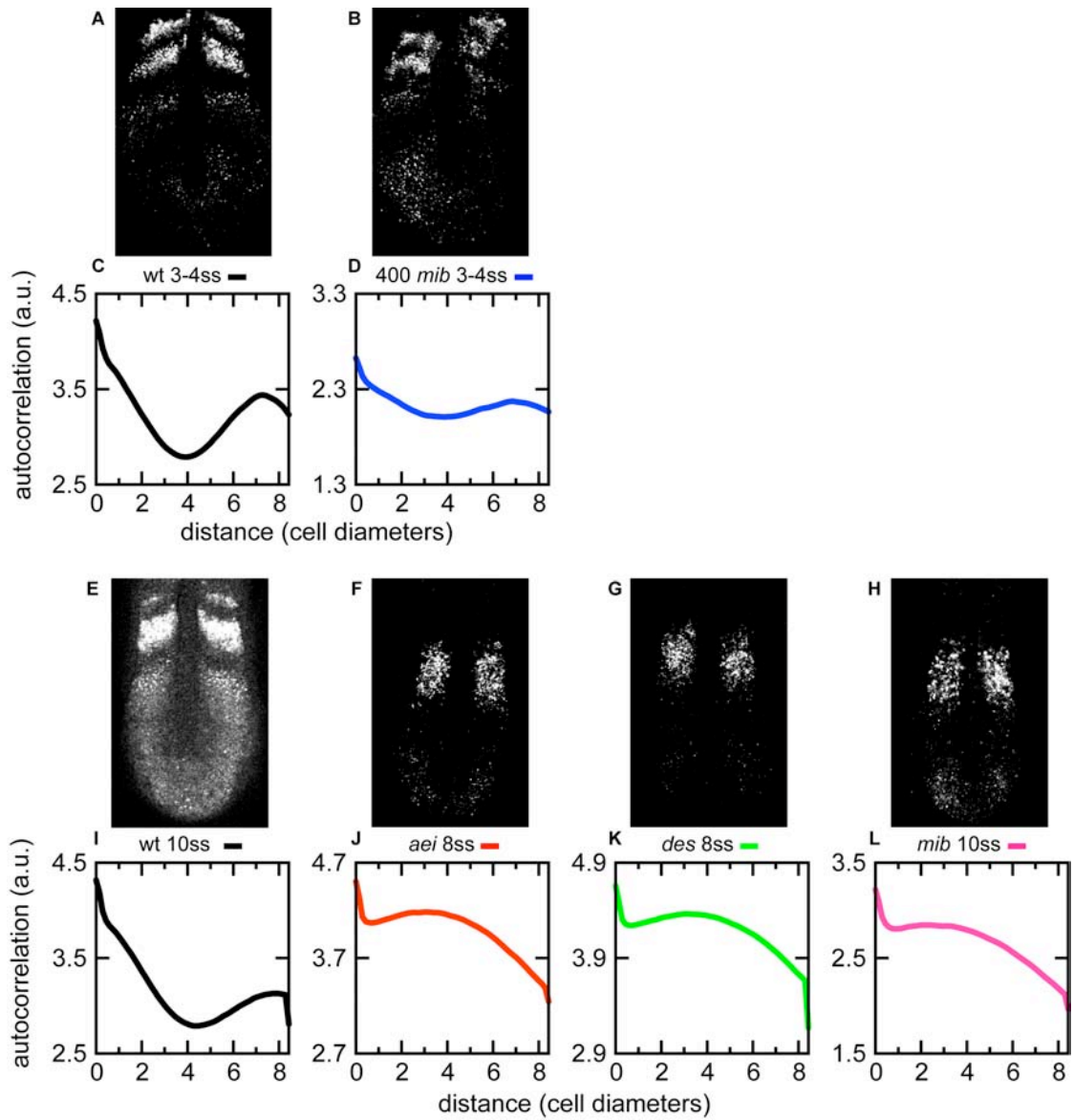

**Figure S7**. *Cyclic gene expression patterns resulting from elevated* **Mib** *levels are distinct from loss of coupling mutants.*

Spatial organization of oscillating *deltaC* expression in the PSM of representative embryos shown for **(A)** wildtype at 3-4 ss, **(B)** 400 pg *mib* mRNA at 3-4 ss, **(E)** wildtype at 10 ss, and the loss of coupling mutants **(F)** *after eight (aei/deltaD)* at 8 ss, **(G)** *deadly seven (des/notch1a)* at 8 ss, and **(H)** *mind bomb (mib)* at 10 ss. Corresponding average autocorrelation functions of oscillatory gene expression patterns in the anterior PSM of (**C**) wildtype at 3-4 ss (n=15), (**D**) Mib over-expression (n=8), (**I**) wildtype at 10 ss (n=13), (**J**) *aei/deltaD* (n=38), (**K**) *des/notch1a* (n=36), (**L**) *mib* (n=26). The arbitrary units in the correlation axis are multiplied by $10^{-3}$. The left/right asymmetry of the pattern in Fig. S7B reflects the asymmetric distribution of the injected mRNA. Autocorrelation functions were always calculated from the most strongly affected side.

## 1.5 Supplemental movies

**Movie S1. Time evolution of the analytical solution of the delayed coupling theory.**
The delayed coupling theory predicts that reduced coupling results in a larger collective period and corresponding changes to the wavelength of spatial patterns. The movie shows the time evolution of the sine of the phase of the oscillations (light-dark scale goes from -1 to 1) in a wildtype zebrafish PSM (top) and a PSM with disrupted coupling (bottom) as described by the analytical solution of the continuum approximation of the delayed coupling theory [7] using parameters determined in this work (see Supplemental Experimental Procedures 2.2.4 and 2.2.6), and illustrated in Figure 1C of the main manuscript. Anterior is to the left and posterior to the right. The blue oscillating pattern on the right represents cyclic gene expression in the PSM, the grey arrested pattern on the left determines the segment length over which somites will form. Noise is not included, hence no desynchronization is observed in the uncoupled system. The period of the individual oscillators is $T_A = 28.5$ min, the coupling strength is $\varepsilon = 0.07$ min$^{-1}$ (top) and $\varepsilon = 0$ (bottom), the time delay in the coupling is set to $\tau = 20.75$ min, the decay length of the frequency of the oscillators along the PSM is $\sigma = 27$ cell diameters, the length of the PSM from the arrest front to posterior end of the notochord is $L = 39$ cell diameters, a 14 cell diameter long tailbud with homogeneous phase occupies the posteriormost part of the PSM, and the velocity of the arrest front is $v = 0.249$ cell diameters/min. With these parameters the collective period is $T_C = 23.5$ min for the wildtype (top) and $T_C = 28.5$ min for the PSM with disrupted coupling (bottom). This leads to a longer wavelength of the oscillating pattern (right, blue) and of the arrested segment length (left, black) in the disrupted system (bottom).

**Movie S2. Anterior trunk somitogenesis in wildtype and *mind bomb* mutant zebrafish embryos.**
Time-lapse movies [8, 9] of wildtype and *mind bomb* embryos during early trunk somitogenesis, corresponding to the upper panels shown in Figure 2B of the main text. Dorsal view with anterior to the top. Starting at the 1 somite stage, the wildtype embryo completes six somite boundaries within 100 min, but the *mind bomb* embryo completes only 5.

**Movie S3. Axial extension during trunk somitogenesis in wildtype and *mind bomb* mutant zebrafish embryos.**
Time-lapse movies [8, 9] of wildtype and *mind bomb* mutant embryos during trunk somitogenesis, corresponding to the upper panels shown in Figure S2A. Lateral view with anterior to the left. Starting from the 3 somite stage, the wildtype embryo completes eight somites within 180 min, whereas the *mind bomb* embryo completes only seven. Somite boundaries are not visible in the posterior region of the *mind bomb* trunk because of the desynchronization of the segmentation clock at later times. Importantly, the length of the embryonic axis in the *mind bomb* embryo is similar to that observed in its wildtype sibling throughout the movie. The superimposed measurement lines shown in some movie frames contribute to the data in Figure S2E.

## 2. Supplemental Experimental Procedures

The material in the Supplemental Experimental Procedures expands on the Experimental Procedures in the main text to give a comprehensive description of the methods we introduced and used in this work. In order for the Supplemental Experimental Procedures to be as complete and self-contained as possible, there is necessarily some repetition of the material from the main text.

### 2.1 Embryology, microscopy and molecular biology

#### 2.1.1 Fish care and mutant stocks

Zebrafish *Danio rerio* were raised and kept under standard laboratory conditions [13]. Embryos obtained from natural spawnings were staged as described [14]. Staging by somite number refers to populations of wildtype/heterozygous or control treated embryos. For obtaining closely staged embryos, two parental fish of relevant genotypes were placed in a mesh-bottomed breeding box and allowed to produce embryos for approximately 15-20 min before harvesting. Mutant alleles used were *aei* $^{tr233\ [15]}$, *bea* $^{tm98\ [16]}$, *des* $^{p37a\ [17]}$, *mib* $^{ta52b}$ [18], *ace* $^{ti282a}$ [19], *mbl* $^{tm013}$ [20] and *nof* $^{u11}$ [21].

#### 2.1.2 DAPT treatment

DAPT has been shown to block proteolytic cleavage of the Notch cytoplasmic domain [22], and can be used to reduce Delta-Notch coupling in a dose-dependent manner [23]. For simultaneous time-lapse imaging of embryos treated with different DAPT concentrations, mold arrays were produced from Silicone elastomer Sylgard 184 (Dow Corning) for holding embryos during imaging, which allowed precise control of DAPT concentration. A Petri dish was subdivided into four chambers by separating plastic plates, and mold arrays were placed into each of the four chambers. E3 containing the desired amount of DAPT or the highest utilized concentration of DMSO carrier was put into each chamber, and embryos were transferred to the Petri dish before shield stage. PTU was added to a final concentration of 0.03% at around 24 hpf, and embryos were fixed at 30-36 hpf for ALD determination to verify the effective action of DAPT [23].

For determination of embryo length and somite length in embryos treated with saturating amounts of DAPT, Petri dishes were subdivided into two chambers, and agarose mold arrays were created in each chamber. A volume of E3 medium equal to the volume of E3/agarose used for the mold array, and containing 100 μM DAPT or the corresponding amount of DMSO carrier, was added to each chamber. Concentrations in E3/agarose and E3 medium were allowed to equilibrate for at least one hour, and embryos younger than shield stage were then transferred to the Petri dish.

For determination of segment length, DAPT treatment was carried out in small Petri dishes or 6-well plates. Embryos were transferred to the Petri dish before shield stage, allowed to develop until the desired stage and fixed for *in situ* hybridization.

*2.1.3 Statistical analysis*

Statistical significance within all relevant data sets concerning somitogenesis period, somite length, segment length at the arrest front, the position of the arrest front, and axial elongation was assessed using Student's t-test, two-sided, unequal variance, in *Microsoft Excel*. All statistical display items were compiled using *OriginLab Origin*.

*2.1.4 Determination of somitogenesis period*

Somitogenesis period was determined as previously described [8, 9]. Briefly, forty closely staged embryos were time-lapsed for 6 h in 5 min time intervals starting around 10 hpf. Somitogenesis period of individual embryos was manually determined from the resulting movies. Movie analysis was performed without knowledge of the embryo's genotype, and results were consistent between three independent observers. All experiments but one were carried out at temperatures between 27 and 29°C, and one *mib* experiment was carried out at 30.1°C. Dorsal imaging allowed monitoring of the formation of somites one to eight, whereas lateral imaging was used to image somitogenesis from somite four to somite 20. Both approaches were used in all experimental conditions to characterize somitogenesis period of somites one to twenty, and results from both approaches were in good agreement.

Somitogenesis period was determined in clutches obtained from heterozygous incrosses containing wildtype, heterozygous and homozygous mutants embryos. Homozygous Delta-Notch [4, 5] and *ace/fgf8* (our unpublished observations) mutant embryos display defects in the formation of posterior somites; see Table 1 for their respective anterior limits of defects (ALDs). Therefore, somitogenesis period was determined only in the anterior, correctly forming somites in homozygous mutant embryos, and somitogenesis periods obtained from corresponding somites in their wildtype siblings were used for comparison. In embryos treated with various DAPT concentrations, somitogenesis period was determined from the first six somites irrespective of the ALD induced by the respective DAPT concentration to ensure comparability of all measured values.

Where necessary, embryos were PCR genotyped after time-lapse imaging. Primer sequences are available from the authors upon request. Somitogenesis periods of individual embryos were grouped according to genotype, and all values were normalized to the mean wildtype somitogenesis period, enabling pooling of results from different experiments.

Homozygous *ace/fgf8, mbl/axin1* and *no fin/raldh2* mutant embryos were identified by inspection of morphological defects apparent in later development [19-21]. The distribution of period values in the wildtype/heterozygous population was unimodal, thereby excluding that differences between wildtype/heterozygous and homozygous populations were obscured by pooling of wildtype and heterozygous embryos. All values were normalized to the mean somitogenesis period of the wildtype/heterozygous embryo population for these mutants.

### 2.1.5 In situ *hybridization and image acquisition*

*In situ* hybridization was carried out as previously described [24]. Probes used were *deltaC* [24], *isl1* [25], *mespa* and *mespb* [26], *spt* [2], *gata1* [3], *egr2b* [1] and *cb1045* [23]. Where necessary, embryos were flat-mounted and photographed using either an Olympus SZX12 or a Zeiss Axioskop 2 using QCapture software. Images were processed using ImageJ and Adobe Photoshop software. ALD determination was carried out as previously described [23]. Hybridization with *mespa* probe was used to distinguish Delta-Notch mutant embryos from their wildtype or heterozygous siblings. In *mib* embryos, both *mespa* and *isl1* could be used for this purpose.

### 2.1.6 Determination of somite length

Agarose molds were created in a Petri dish as described [9] and covered with E3 medium. Agarose molds were overlaid with methylcellulose, and embryos in their chorions at approximately 5 somite stage were left to sink into the molds whilst being supported by the agarose, which facilitated mounting. Embryos were dechorionated and photographed on an Olympus SZX12 using QCapture software. If necessary, embryos were left to develop until the next day, when homozygous mutants could be identified by inspection of somitogenesis defects.

Lengths of somites two to five were determined from the photographs using Adobe Photoshop software. The length of somite one was not measured because its anterior border was not clearly distinguishable in most embryos. Measurements were performed without knowledge of the embryo's genotype, and results were consistent between three independent observers. Individual embryos were assigned to wildtype/heterozygous or homozygous populations, and all measured values for one somite were normalized to the mean of the wildtype/heterozygous population, enabling comparison of results from different experiments.

### 2.1.7 Segment length and wavefront velocity determination

Distance between stripes of *mespb* expression in the anterior PSM as a measure of segment length, and distance of the posterior border of the *mespb* expression domain from the posterior tip of the embryo as a measure of wavefront position, were assessed using Adobe Photoshop software. Measurements were performed without knowledge of the embryo's genotype, and results were consistent between two independent observers. Genotyping of embryos was carried out by double hybridization of *mespb* with *isl1* for *mib* mutants or by PCR genotyping for all other mutants. Individual embryos were assigned to wildtype/heterozygous or homozygous populations, and all measured values were normalized to the mean of the wildtype/heterozygous population, enabling comparison of results from different experiments.

Embryo length was determined from time-lapse movies of laterally mounted embryos by tracing embryo length using the Segmented Line tool in ImageJ (see Figure S2 for details), and by subsequently determining the length of this line using the Measure tool. Slightly different developmental stages of embryos at the beginning of time lapse imaging in different experiments were accounted for by normalizing elapsed recording time to the time of formation of the fourth somite.

### 2.1.8 Cyclic gene expression stripe wavelength measurements and parameter fitting

The oscillatory properties of the cells in the PSM can be inferred from the patterns of gene expression collectively generated at the tissue level using the delayed coupling theory. For this purpose, we measured the anterior-posterior wavelength of each stripe of gene expression (the stripe wavelength, $\lambda$) and the corresponding position $x$ of the stripe within the PSM. This method assumes that cells at equivalent positions in successive gene expression stripes have equivalent phase [7, 27]. The stripe wavelength $\lambda$ of *deltaC* expression was assessed in flat-mounted embryos by measuring the distance between the anterior borders of successive gene expression stripes immediately adjacent to the notochord with the measure tool in Adobe Photoshop (see Fig. 4A in main text). Data points were collected only from stripes with well-defined anterior borders, therefore most data points collected are from stripes in the anterior half of the PSM. The position $x$ of each stripe wavelength is it's anterior-posterior midpoint within the PSM (see Fig. 4A in main text, and 2.2.4 below), which we measured as the distance between the centre of a wave (defined as a stripe and its posteriorly situated interstripe) of gene expression and the arrest front of cyclic gene expression in the anterior PSM.

We carried out this experiment in *mib* embryos because *mib* displayed pronounced changes of comparable magnitude in somitogenesis period and segment length. The *deltaC* gene was chosen because the mRNA expression levels of two other cyclic genes, *her1* and *her7*, are downregulated in Delta-Notch mutants [24], which prevented meaningful stripe wavelength measurements from expression patterns of these genes for the relevant experimental conditions. Genotyping of *mib* embryos was carried out using double *in situ* hybridization of *deltaC* with the *isl1* probe.

A subset of embryos was stained with Hoechst 33342 (Sigma-Aldrich) to visualize nuclei, and sections through the PSM at the level of the notochord were taken on a Leica TCS SP2 confocal system. Number of nuclei in a rectangular region of interest comprising the anterior half of the PSM was determined in ImageJ using the Nucleus Counter plugin. We did not detect any statistically significant differences in the number of nuclei in wildtype/heterozygous embryos as compared to their mutant siblings at approximately the 4 somite stage ($p > 0.01$, n (wt) = n (*mib*) = 36 embryos from four independent trials). Measurements were normalized to the mean value of the wildtype/heterozygous population in each experiment. Means values were calculated with 95% confidence intervals: 100 ± 2 % for both wildtype/heterozygous and *mib* populations. This finding indicated that the observed increase in stripe wavelength were not due to a change in cell density in the anterior PSM of *mib* mutants.

Estimation of the parameters from the fit of the DCT to the stripe wavelength data is described in section 2.2.4, and bootstrap evaluation of the parameter significance is describe in section 2.2.5.

### 2.1.10 Plasmids, mRNA synthesis and embryo microinjection

Zebrafish *mind bomb* was contained in the pEGFP3 vector, where eGFP had been replaced with 5 myc tags (kindly provided by Wen Biao Chen, Oregon Health and Science University, Portland, Oregon, USA). *pCS2+* containing eGFP was kindly provided by Magdalena Strzelecka (MPI-CBG, Dresden, Germany). Synthesis of mRNA was carried out as previously described (Oates and Ho 2002), and delivered to one-cell stage embryos using our quantitative injection protocol (Riedel-Kruse et al. 2007). After injection, embryos were left to develop until they reached the 3-4 somite stage when they were fixed for *in situ* hybridization. The effects were usually restricted to the left- or right-hand side of the embryo due to unequal distribution of mRNA, as expected for zebrafish microinjection and followed by tracer GFP mRNA, and we analyzed the side with the stronger defects.

## 2.2    Theoretical methods and parameter estimation

### 2.2.1    Overview of the delayed coupling theory (DCT)

The delayed coupling theory (DCT) describes the PSM as an array of coupled phase oscillators [7], coarse-graining the various underlying molecular processes into effective variables and parameters. The state of oscillator *i* at time *t* is described by a phase variable $\theta_i(t)$ whose dynamics is given by

$$\frac{d\theta_i(t)}{dt} = \omega_i(t) + \frac{\varepsilon_i(t)}{N_i} \sum_k \sin[\theta_k(t-\tau) - \theta_i(t)] + \eta\zeta_i(t) \ . \qquad \text{(main text Eq. 2) (S1)}$$

The frequency profile $\omega_i(t)$ describes the slowing down of oscillators as they get closer to the arrest front, where oscillations stop. Oscillators are coupled to their neighbors with a coupling strength $\varepsilon_i(t)$, and a coupling delay $\tau$. Oscillator *i* is coupled to its $N_i$ nearest neighbors, labeled by *k*. The zero average un-correlated random term $\zeta_i$ represents different noise sources with noise strength $\eta$. All the other expressions of the delayed coupling theory that we use in this work derive from our previous analysis of Eq. (S1) [7].

The collective and autonomous frequencies in our phase description correspond directly to the observable frequency of oscillations of coupled and uncoupled cells, respectively. A fundamental feature of reduced models is that their effective parameters describe the combination of several different microscopic processes that together give rise to the observed macroscopic behavior. Thus, the coupling strength and coupling delay in the DCT are not necessarily related in a simple way to single molecular events. The coupling strength $\varepsilon$ in our model is not precisely the number of molecules participating per unit time in a coupling event, although it will be a function of this number. In the same way, the coupling delay in the phase description arises from a combination of events, being a function of the typical time that it takes for a Delta molecule to signal (transcription, translation, and trafficking etc.). Thus, the value that we find for the time delay in the phase description cannot currently be directly compared to previous estimations of signaling delays, for example using molecular models.

Even without knowledge of the functions relating the molecular processes to the coarse-grained parameters, the values of these parameters obtained by fitting to the data are invaluable for interpreting experimental perturbations. Firstly, a quantitative change to only one parameter, for example coupling strength, excludes that some other process in the system is altered. Secondly, a quantitative change in coupling strength would be an indication that the number of active signaling molecules per unit time was altered. Likewise a change in the coupling delay would indicate an alteration to the transcription, translation, and trafficking times involved in producing an active Delta signal. From the identity of the affected parameters, the specific molecular players or pathways used to perturb the system can be assigned to functions at a macroscopic, or tissue level. Further, perturbations for which the specific molecules are not known can be assigned tissue-level function, thereby suggesting which molecular pathways may be involved. Finally, from the values of the parameters, the relative strength of the perturbation can be determined.

### 2.2.2 Stability of the solutions of main text Eq. (1)

Eq. (1) from the main text provides a relationship between the collective period $T_C$, the autonomous period $T_A$, the coupling delay $\tau$ and the coupling strength $\varepsilon$:

$$\frac{2\pi}{T_C} = \frac{2\pi}{T_A} - \varepsilon \sin\left(\frac{2\pi}{T_C}\tau\right). \qquad \text{(main text Eq. 1) (S2)}$$

A periodically oscillating solution to the dynamic equation of the delayed coupling theory obeys this relation. Such a solution is stable if $\varepsilon \cos(2\pi \tau/T_C) > 0$ [7, 30]. Fig. 5B in the main text shows the stable branches of solutions as solid lines and the unstable branches as dashed lines.

### 2.2.3 Phase offset

In the segmentation clock, the main role of coupling is to synchronize the genetic oscillations of neighboring cells. For this reason we have chosen in our theory an attractive coupling that minimizes the delayed phase difference of neighboring oscillators. In other systems, coupling can play alternative roles. For example, in other instances of Delta/Notch signaling, as in lateral inhibition, the coupling can be repulsive [42]. Also in neuroscience, coupling between neurons can be inhibitory. In these cases, the effect of the coupling is to push neighboring cells towards a non-vanishing phase difference. This situation can be described in a theory of phase oscillators by introducing a phase offset $\alpha$ in the coupling. There is a close relationship between the effects of a phase offset and that of delayed coupling. In the presence of coupling delays, the coupling can be tuned to behave as attractive or repulsive, depending on the relative value of the delay to the period of oscillations and the phase offset.

For simplicity, in Eq. (S1) we chose not to include a phase offset in the coupling. A more general description [43] could include a phase offset $\alpha$ in the argument of the coupling function:

$$\frac{d\theta_i(t)}{dt} = \omega_i(t) + \frac{\varepsilon_i(t)}{N_i} \sum_k \sin[\theta_k(t-\tau) - \theta_i(t) - \alpha] + \eta\zeta_i(t) \,. \tag{S3}$$

Note that the coupling function $\sin[\theta_k(t-\tau) - \theta_i(t) - \alpha]$ vanishes when the delayed phase difference matches the value of this phase offset. The phases of neighboring oscillators get attracted to match this phase difference, *e.g.* in a system of two coupled oscillators, without delayed coupling, these would synchronize with a constant phase difference of $\alpha$.

Due to the properties of the coupling function, $\sin[\theta_k(t-\tau) - \theta_i(t) \pm \pi] = -\sin[\theta_k(t-\tau) - \theta_i(t)]$, a phase offset of $\alpha = \pm\pi$ introduces a change of sign that turns an attractive coupling into a repulsive coupling and *vice versa*. Coupling delays can further change the nature of the interaction. For example, an attractive coupling $\varepsilon > 0$ and $\alpha = 0$ for vanishing delays becomes repulsive when $\tau = T_C/2$, and a repulsive coupling $\varepsilon < 0$ and $\alpha = 0$ becomes attractive when $\tau = T_C/2$.

For given values of the coupling strength $\varepsilon$ and phase offset $\alpha$, the uniformly rotating steady state solution satisfies the relation

$$\frac{2\pi}{T_C} = \frac{2\pi}{T_A} - \varepsilon \sin\left(\frac{2\pi}{T_C}\tau + \alpha\right), \tag{S4}$$

which reduces to Eq. (S1) for $\alpha = 0$. The corresponding stability condition is now $\varepsilon \cos(2\pi\tau/T_C + \alpha) > 0$. When this stability condition is fulfilled, the nature of the coupling is attractive and results in locally synchronized uniformly rotating solutions satisfying Eq. (S4). Outside the ranges defined by the stability condition, the coupling becomes repulsive and can result in patterns with a phase difference between neighboring oscillators, as observed for example in lateral inhibition [42, 44].

The consequence of introducing the phase offset $\alpha$ to the curves displayed in main text Fig. 5B is a shift, in such a way that for $\tau = 0$ the value of the collective period $T_C$ is different from the value of the autonomous period $T_A$, see Fig. S8A. For fixed values of the coupling $\varepsilon$, autonomous period $T_A$, and collective period $T_C$, the value of the time delay $\tau$ is given in terms of the phase offset $\alpha$ by

$$\tau(\alpha) = \frac{T_C}{2\pi}\left(\arcsin\left(\frac{2\pi}{\varepsilon}\left(\frac{1}{T_A} - \frac{1}{T_C}\right)\right) - \alpha\right), \tag{S5}$$

as illustrated in Fig. S8B. Equation (S5) defines the family of possible pairs of values of $\tau$ and $\alpha$ that yield equivalent steady state solutions, representing systems with the same collective period $T_C$ of the synchronized oscillations, for given values of the parameters $\varepsilon$ and $T_A$.

The steady state described by Eq. (S4) is invariant under the transformation

$$\alpha \to \alpha'$$
$$\tau \to \tau' = \tau + (\alpha - \alpha')\frac{T_C}{2\pi}$$

which leaves the argument of the sine in Eq. (S4) unchanged. For example, a steady state with an arbitrary phase offset $\alpha$ can be mapped to its equivalent steady state with $\alpha' = 0$ by defining a new effective delay $\tau' = \tau + \alpha\, T_C/2\pi$. Therefore, in this work we use the convention $\alpha = 0$ for simplicity, without loss of generality.

In steady state, one cannot independently determine $\tau$ and $\alpha$, but only the combination $2\pi\tau/T_C - \alpha$. Because this transformation is not a symmetry of Eq. (S3), transients and out of steady state conditions could be used to separate $\tau$ and $\alpha$ value pairs.

In summary, consideration of phase offset connects the particular simplified version we have used to understand the experimental data to the more general cases. In this context, our choice of a positive coupling strength $\varepsilon$ does not imply a particular biological mechanism. Because of the symmetries of the solution, a negative coupling strength could also be used to fit the data without changing the main conclusion that coupling delays are responsible for the change in collective period.

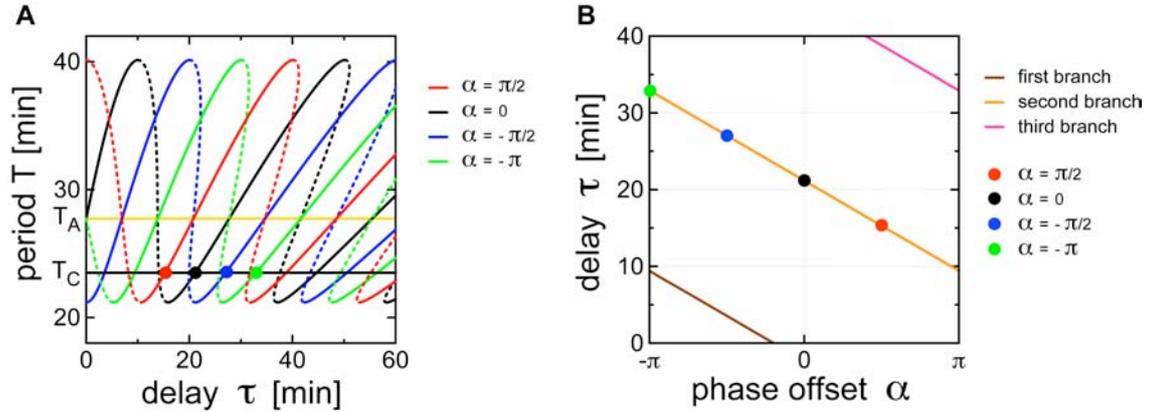

**Figure S8. Effects of a phase offset on the uniformly rotating steady state solutions.**
**(A)** Diagram of solutions to Eq. (S4). Collective period $T_C$ of steady state solution vs. coupling delay $\tau$ for different values of the phase offset $\alpha$, as indicated. Stable and unstable solutions: solid and dashed lines, respectively. The dots display the positions in the diagram of equivalent steady state solutions with the same collective period $T_C$. Parameters as in main text Fig. 5, $\varepsilon = 0.07$ min$^{-1}$. **(B)** Relation between $\tau$ and $\alpha$ for equivalent steady state solutions, Eq. (S5).

*2.2.4 Parameter estimation from cyclic gene expression stripe wavelength fitting*

The delayed coupling theory (DCT) describes the appearance of characteristic changes to the wavelength of oscillating gene expression patterns in the PSM of a mutant with altered period. Cyclic gene expression patterns are related to the frequency profile in our theory. To fit cyclic gene expression patterns, it is convenient to use the continuum approximation to main text Eq. (2) (Eq. (S1)) in the reference frame co-moving with the PSM [7]. We assume that the intrinsic frequency at a distance $x$ from the arrest front is $\omega(x) = \omega_A (1 - e^{-x/\sigma})/(1 - e^{-L/\sigma})$, where $x$ is a continuous variable, $\omega_A = 2\pi/T_A$ is the frequency of the individual oscillators in the posterior PSM, $\sigma$ is the decay length of the frequency profile, measuring the characteristic distance over which the frequency decreases from high to low values, and $L$ is the distance between the posterior end of the notochord and the arrest front of the oscillations. More posteriorly, the frequency of the oscillation is constant. The phenomenological function $\omega(x)$ for the frequency profile compares well with experimental data and is consistent with the shape determined in previous work [7, 27]. The positions $x$ and wavelengths $\lambda$ of the gene expression stripes are related by the expression [7]:

$$x = \sigma \ln\left[\frac{2\sigma \sinh(\lambda/2\sigma)}{S(1 - e^{-L/\sigma}) + \lambda e^{-L/\sigma}}\right], \qquad \text{(main text Eq. 3) (S6)}$$

where $S$ is the arrested segment length. Note that the patterns of expression are not sensitive to spatial variation in coupling strength or delay. This is because the first order approximation of the relation of wavelength and position [7] that we use to fit the wave patterns does not depend explicitly on the coupling strength or the coupling delays (Eq. (S6)). The effect of a hypothetical spatial dependence on coupling strength or coupling delay would be present as a higher order correction to Eq. (S6). that can be safely neglected due to the fact that coupling strength is the smallest parameter in the model. There is an implicit dependence on these parameters through the segment length $S$, which itself depends on the collective period, which in turn depends on coupling strength and coupling delays through Eq. (S2). Here the contribution comes only from the values at the posterior of the PSM, where the collective period is determined, and therefore variations along the PSM are not relevant. Thus, for simplicity we assumed the coupling strength and coupling delay to be constant across the PSM.

The position of the arrest front is difficult to determine from *in situ* experiments, since it cannot be known if the gene expression in the anterior-most part of the PSM is still oscillating or already arrested. We measured the distances between the anterior borders of the stripes of gene expression and the posterior end of the PSM. The difference between the measurements of two consecutive borders is the wavelength $\lambda$ of the gene expression stripe, and the mean is the position of the stripe relative to the end of the PSM. However, the relevant boundary in the delayed coupling theory is not the posterior end of the PSM, but the posterior end of the notochord, where the oscillators start their slowing down. We measured the size of the tailbud region between the end of the notochord and the posterior end of the PSM and found that it scales with good accuracy as $2L/5$. This scaling of the tailbud length was checked successfully for consistency at the end of our fitting procedure. We denote the position of a stripe relative to the end of the PSM by $\Delta$: it is related to the position $x$ of the stripe with respect to the arrest front by $\Delta = 7L/5 - x$. Substituting $x$ by $\Delta$ we can rewrite Eq. (S6) as:

$$\Delta = \frac{7L}{5} - \sigma \ln\left[\frac{2\sigma \sinh(\lambda/2\sigma)}{S(1 - e^{-L/\sigma}) + \lambda e^{-L/\sigma}}\right]. \tag{S7}$$

The data set obtained from *deltaC* stripe wavelength measurements in wildtype and *mib* was used for determination of segment size as well as the relative slowing down of oscillations in the PSM along a frequency profile by fitting to Eq. (S7). There are three unknown parameters in Eq. (S7): the length of the considered region of the PSM $L$, the decay length of the frequency profile $\sigma$, and the segment length $S$. Our data does not constrain a direct fit to Eq. (S7) with three free parameters. To better constrain our fit, we first measured $S$ directly from wildtype embryos as the average of the width of the anterior-most stripe of *deltaC* expression. Using this value of $S$ in Eq. (S7), we fitted the wildtype data with only $L$ and $\sigma$ as free parameters. This fit is robust and yields well defined values for $L$ and $\sigma$. Since we do not observe changes in the size of the PSM between wildtype and *mib* embryos (Figure S3), the value of $L$ for *mib* embryos is assumed to be the same as determined for wildtype. As a consistency check, we used the value of $L$ as a fixed parameter in Eq. (S7) and fitted the wildtype and *mib* data, obtaining from these fits values of $\sigma$ and $S$ for both wildtype and *mib* conditions. The values of $S$ and $\sigma$ obtained from this second wildtype fit are in agreement with the value of $S$ used as a fixed parameter in the first wildtype fit and the value of $\sigma$ obtained from it, showing the consistency of our approach. The value of $S$ obtained from the *mib* fit is consistent with the direct measurement of the anterior-most stripes. Since both $x$ (or $\Delta$) and $\lambda$ come from measurements of the same observable, the position of stripe borders, they are both affected with the same experimental errors. To take this into account, the fit of Eq. (S7) to the data has been done with a total least square method [28]: minimizing the sum of the squared distance of the fitted curve to both $x$ (or $\Delta$) and $\lambda$, using custom designed software available upon request.

With our procedure we have determined the segment size and the frequency profile of the slowing down of the genetic oscillations directly from gene expression data, and also the position of the arrest front, given by the parameter $L$. The results of the fits are shown in Fig. 4B,C of the main text. To enable comparison with species other than zebrafish, we show the data normalized by the calculated length $L$ and give as results the normalized parameters $\rho = \sigma/L$ and $s = S/L$. In these normalized units, the distance between the end of the notochord and the arrest front is 1 and $\rho$ and $s$ can be directly interpreted as fractions of this distance. As we did not find evidence of altered arrest front velocity, position of the arrest front or PSM length in *mib* embryos as compared to their wildtype siblings (Figures. S2, S3), we used the value of $S$ obtained from the fit as a measure of oscillator period according to $S = vT_C$ [6, 7].

### 2.2.5 Bootstrap evaluation of parameter significance

The errors of the fitted parameters $\rho$ and $s$ were determined using a statistical bootstrap procedure [29] (main text Fig. 4C). A set of $N$ data points is selected randomly from the initial $N$ data points. This bootstrapped set, where some of the data points do not appear and others are repeated, is fitted using Eq. (S7) with $L$ fixed as discussed in the previous section, yielding new values for the parameters $\rho$ and *s*. If this process is repeated *M* times, we obtain a set of *M* estimated values of $\rho$ and *s*. From this set a mean and a standard deviation of $\rho$ and *s* can be calculated. For large enough *M* the calculated means and standard deviations converge and do not change by further increasing the number *M* of bootstrapped samples. From the standard deviations obtained by this procedure, we calculated the confidence intervals of $\rho$ and *s* given in the main text.

### *2.2.6 Fit of somitogenesis period vs. DAPT treatment concentration*

Eq. (S2) yields possible values of the collective period $T_C$ of the segmentation clock for given autonomous period $T_A$, coupling strength $\varepsilon$, and coupling delay $\tau$. These values are represented as sets of curves in main text Fig. 5B. The sets of curves appear complex, possessing regions of instability and multistability. Depending on whether the coupling delay is relatively short or long compared to the autonomous period, the system's collective period can either be larger or smaller, respectively, than the autonomous period, which lies along the $\varepsilon = 0$ curve. As the coupling delay grows larger than the autonomous period, the regions of the solution space that harbor unique and stable solutions get smaller and multistability comes to dominate. We assume that only uniquely stable solutions will be evolutionarily selected, as they imply robust function. Nevertheless, it is a prediction of the theory that multistability can exist for long coupling delays – the technical means of introducing these into the system will be of great interest.

An overview of the measurements of segmentation variables in the various experimental conditions (Table 1) reveals that the direction and order of magnitude of the observed changes in period, segment length at the arrest front, and somite length were consistent in all experimental conditions. However, some variance was observed, possibly associated with additional roles for the genes under investigation in processes downstream of the segmentation clock such as: how the arrest wavefront stops the oscillators, the generation of segment polarity, and the formation of the epithelial somite boundaries. These aspects of somitogenesis are outside the scope of the DCT. Somitogenesis period (main text Fig. 2, Fig. 5A) is known with much greater accuracy and precision in our work than segment length (main text Fig. 3), and because it matches closely with the collective period predicted from oscillating gene expression stripe wavelength measurements (main text Fig. 4), we use it for parameter calculations with the DCT.

To localize the DCT solution corresponding to the wildtype zebrafish segmentation clock in the space of Figure 5B, parameters $\varepsilon$, $T_A$, and $\tau$ must be estimated from experimental data. We first make the simplifying assumption based on previous studies [23, 31] that autonomous period is essentially unaffected in DAPT-treated embryos during the time window of interest. Previous work [23] also allows the reasonable assumption that saturating DAPT disrupts coupling completely. Following [23] we also assume here that DAPT suppresses Notch signaling according to Michaelis-Menten kinetics, [Notch]~$1/(1+n/n_0)$, where $n_0$ is the DAPT concentration that halves Notch signaling. Assuming a linear relation between coupling strength and the level of Notch signaling we obtain $\varepsilon(n) = B/(1+n/n_0)$, where $B$ is a constant related to the wildtype level of coupling receptor (Notch) expression. Fig. 5A of the main text shows that the collective period increases with DAPT treatment. As shown in Fig. 5B, the prediction of the DCT is that this increase in the collective period occurs only if there is a time delay in the coupling different from zero. A short delay (branch 1 in Fig. 5B) corresponds to a shorter collective period with coupling disruption, hence the delay in the system has to be long enough to place the solution in branch 2 or 3, where solutions with increased collective period are found.

The value of $B$ is bounded from below because according to Eq. (S2) there is a minimum value of the coupling strength $B_{min} \approx 0.043$ min$^{-1}$ that allows for the observed variation in the collective period when the coupling is completely disrupted. The wildtype coupling strength $B$ should be larger than this value. For this value of the coupling, there is multistability from the third branch onwards. As explained above, we assume that the parameters of the zebrafish segmentation clock lie in a region where no multistability is possible. This confines the time delay to branch 2 in main text Fig. 5B. Under this assumption, the collective period for a DAPT concentration $n$ can be approximated from Eq. (S2) by:

$$T(n) = T_A \frac{1+\varepsilon(n)\tau}{1+\varepsilon(n)T_A} = T_A \frac{1+n/n_0+B\tau}{1+n/n_0+BT_A}. \tag{S8}$$

According to our previous assumption, saturating DAPT disrupts coupling completely, so the autonomous period $T_A$ is the limiting value of $T(n)$ for a saturating treatment with DAPT (n→∞). The untreated wildtype period $T_u$, is the value for zero DAPT treatment, $T_u = T(0)$. Therefore

$$T_u = T_A \frac{1+B\tau}{1+BT_A} \Rightarrow \begin{cases} \tau = T_u - \dfrac{T_A - T_u}{BT_A} \\ B = \dfrac{1}{T_A} \dfrac{T_A - T_u}{T_u - \tau} \end{cases}. \tag{S9}$$

Both $T_u$ and $T_A$ are experimentally well determined quantities shown in Figure 2E and Table I. Eq. (S9) relates the coupling strength $B$ to the coupling delay $\tau$, indicating that only one of them is an independent parameter. Eqs. (S8) and (S9) can be used to write the ratio between the period observed for a given DAPT treatment level and the untreated period. Including a factor 100 to express the result as a percentage gives:

$$\frac{T(n)}{T_u}\% = 100 \frac{nT_A/T_u + Q}{n+Q}, \qquad \text{(main text Eq. 4)} \tag{S10}$$

$$Q = n_0(1+BT_A), \qquad \text{(main text Eq. 5)} \tag{S11}$$

The parameters $n_0$ and $B$ appear only through the combination $Q$ defined in Eq. (S11). Since $T_A$ and $T_u$ are known, the ratio $T(n)/T_u$ is a function of $Q$ alone. For the zebrafish wildtype embryo the collective period at 28°C is $T_u = 23.5 \pm 1$ min [9], and the autonomous period is $T_A = 28 \pm 1$ min (main text Fig. 2E, Table I). The curve shown in Fig. 5A of the main text is a fit using Eq. (S10), made with the Curve Fitting Toolbox of Matlab, weighting the data with the inverse of their variance, and using robust fitting together with the Levenberg–Marquardt algorithm for minimization of the residuals. From this fit we obtain $Q = 7.5 \pm 3.5$ μM.

Although we cannot evaluate $B$ and $n_0$ independently, we can plot the relationship between them

$$B(n_0) = \frac{1}{T_A}\left(\frac{Q}{n_0} - 1\right), \qquad (S12)$$

using the value of $Q$ obtained from the fit, with an error given by

$$\Delta B(n_0) = \sqrt{\Delta Q^2 \frac{1}{(n_0 T_A)^2} + \Delta T_A^2 \left(\frac{1 - Q/n_0}{T_A^2}\right)^2} \qquad (S13)$$

shown in Fig. S9A.

Large values of coupling strength would make the collective period very sensitive to changes in the coupling delay, evidenced by steeper slopes to the solutions in main text Fig. 5B. With a high coupling strength, decreasing delays from the wildtype values would result in the system moving along a large region of branch 2 with significantly shorter collective periods. An upper bound to coupling can therefore be estimated from the experimental observation that the period of somitogenesis did not increase dramatically upon Mib over-expression. Furthermore, a coupling strength higher than $\varepsilon = 0.14$ min$^{-1}$ causes the onset of multistability in branch 2, which would compromise the robustness of the system, and which can be appreciated by the increasing tilt of the curves for larger $\varepsilon$ in Fig. 5B. As mentioned above, the value of $B$ is also bounded from below because according to Eq. (S2) there is a minimum value of the coupling strength $B_{min} \approx 0.043$ min$^{-1}$ that allows for the observed variation in the collective period when the coupling is completely disrupted. These considerations allow us to give a rough estimate to the value and range for the coupling strength $B = 0.07 \pm 0.04$ min$^{-1}$.

Coupling delay $\tau$ can be estimated from Eq. (S9) and our estimates for the other parameters. The value of $\tau$ is not strongly dependent on the value $B$, as illustrated by the curve plotted in Fig. S9B. As mentioned above, the value of $B$ is not sensitive to changes in $n_0$ for a moderate range of the treatment level $n_0$ corresponding to the experimental situation (Fig. S9A). Combining these arguments and propagating the errors from Eq. S9, we can give a relatively accurate estimation for the coupling delay in our theory, $\tau = 21 \pm 2$ min.

Although the value of the coupling strength is not well constrained by the data, this does not affect our conclusions. In the diagram where we plot the wildtype and other experimental conditions (main text Fig. 5C) the absolute values are not critical: the relative position of the different experiments in the diagram will be robust.

Note that we cannot experimentally rule out the possibility that the delay is in a branch $m$ larger than the second in Fig. 5B, in which case the approximate expression for the delay reads:

$$\tau_m = (m-1)T_u - \frac{T_A - T_u}{BT_A}. \tag{S14}$$

Hence, our data supports a time delay different from zero, but without the hypothesis of non-multistable solutions, see above, we cannot discriminate between a set of possible time delays $\tau_m = (m-1)\,23.5 - 2.5$ min, with $m$ an integer greater than zero denoting the branch of the solution.

Previous work using measurements of synchrony decay times and a mean field theory of coupled oscillators without coupling delays [23], estimated a coupling strength of $\varepsilon = 0.07$ min$^{-1}$. Although the value obtained in [23] is in agreement with our estimate, these parameters have slightly different meanings, and caution must be used when comparing these values because of the differences in the theories.

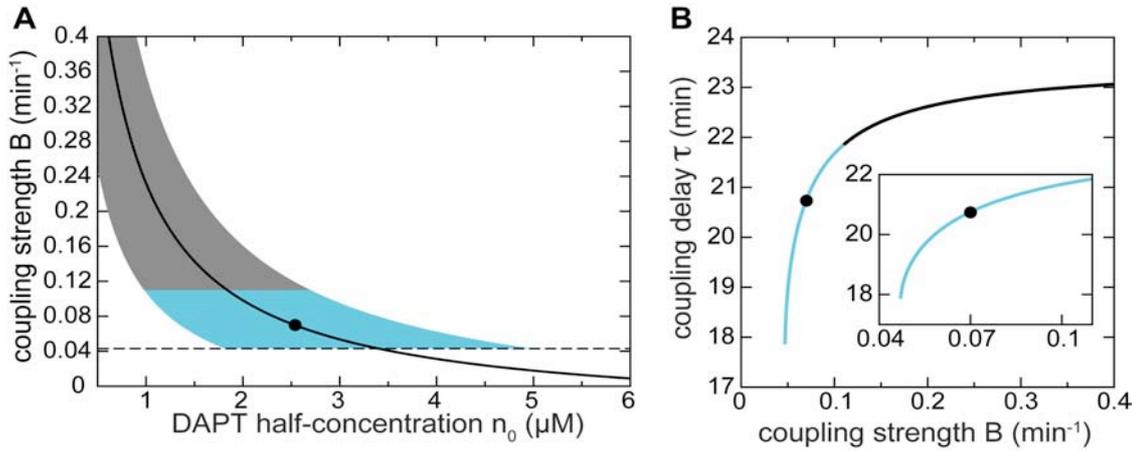

**Figure S9. Estimation of coupling strength and coupling delay.**
(**A**) Continuous line: relation between coupling strength $B$ and the concentration of DAPT that halves the coupling strength $n_0$, as given by Eq. (S9), with $Q = 7.5$ mM, $T_A = 28$ min. Dashed line: lower bound of the coupling strength, $B \sim 0.043$ min$^{-1}$. Black dot: our estimate of the coupling strength $B$. Shaded areas: error of the $B$-$n_0$ relation, given by Eq. (S10) with $\Delta Q = 3.5$ mM and $\Delta T_A = 1$ min. Blue shaded area: error of the $B$-$n_0$ relation for our estimate of $B$. (**B**) Coupling delay $\tau$ as a function of coupling strength $B$, obtained from the second branch of Eq. (1). Black dot: our coupling strength estimation. Blue part of the curve: delay time for the error range of our coupling strength estimation. Inset: zoom of the blue part of the curve, showing that the estimation of $\tau$ is robust to errors in $B$.

*2.2.7 Assumptions for plotting experimental conditions in main text Fig. 5C (collective period vs. coupling delay diagram)*

In order to place the various experimental treatments on the collective period vs. coupling delay state diagram (main text Fig. 5C), we need to propose how the biochemistry of the mutated or inhibited protein can be represented in the DCT [7] by the parameters $T_A$, $\varepsilon$ and/or $\tau$. Previous work has shown that the coupling of oscillating cells is impaired in all Delta-Notch mutants [23, 31, 32] and in DAPT-treated embryos [23, 32] indicating that coupling strength $\varepsilon$ is reduced in all these conditions – this assignment does not need to distinguish between the different biochemical activities of ligands or receptors.

To assign changes in $\tau$, we need to consider the known biochemistry of vertebrate Delta-Notch signaling in more detail [33, 34]. For the zebrafish PSM, a Delta-Notch coupling event is complex [27, 35], beginning with the transcription of a Delta gene, splicing and export of the mRNA and translation of the protein. This is followed by glycosylation, export to the cell surface, and trafficking of the Delta protein through the endocytic pathway, which involves the Mindbomb E3 ubiquitin ligase and is required for efficient Delta-Notch signaling [18, 36-38]. Recent measurements in mammalian cell culture of the endocytic events involved in Delta activation show that these steps take in the order of tens of minutes [39]. After ligand binding to Notch (already present on the target cell surface), γ-secretase cleavage rapidly releases the Notch intracellular domain, which enters the nucleus, binds to the CSL protein (already located at the target promoters), and activates target gene expression, thereby ending the coupling signal. Preparation of the active Delta ligand by the signal-sending cell likely takes much longer than the events following Notch cleavage in the target cell, an assumption used in previous work [11].

Thus, most of the coupling delay $\tau$ is attributable to processes in the signal-sending cell, and the contribution from events in the target cell can be neglected to first approximation. Mutants in *des/notch1a* or application of DAPT would therefore not be expected to alter the coupling delay $\tau$. In contrast, reduction in the levels of active Delta in either *aei/deltaD* mutants (where *bea/deltaC* is still functional), or by dysfunctional Delta trafficking in *mib* [18, 40], should increase the time taken to attain sufficiently high levels of Delta to trigger the target cell's response. In addition to a reduction in coupling strength $\varepsilon$, we therefore expect *aei/deltaD* and *mib* mutants to show an increase in coupling delay $\tau$. These issues are not unique to Delta-Notch signaling or the segmentation clock. Analysis of coupling through other signaling pathways and in other systems will require consideration of the relative times of the different signaling processes.

The rank ordering of the mutants and treatments by collective period change is not the same as the ranking of the severity of segmentation defects, as measured by the onset of segmentation defects, termed the Anterior Limit of Defects (ALD) [5, 23, 24, 41]. This finding seems at first glance counterintuitive. However, the ALD is caused by the decay time for desynchronization of the individual oscillators, which depends primarily on coupling strength $\varepsilon$ [23], whereas collective period $T_C$ is controlled by coupling strength $\varepsilon$ and coupling delays $\tau$. Importantly, the wildtype zebrafish segmentation clock is located below the $\varepsilon = 0$ curve in main text Fig. 5C: from this position, reduction of coupling strength $\varepsilon$ moves the solution curve up, yielding longer collective period $T_C$, e.g. *des/notch1a*, and increase of coupling delay $\tau$ moves the point along the solution curve to the right, also increasing the collective period $T_C$, e.g. *aei/deltaD*. This offers a graphical explanation for why these two mutants with the same onset of segmentation defects have different somitogenesis periods.

### *2.2.8 Movie of the analytical solution's time evolution*

The sine of the phase in the PSM depicted in Movie S1 is obtained from the analytical solution of the delayed coupling theory in its continuum approximation [7]. The phase $\phi$ at a relative position $\xi$ in the PSM (where $\xi = 0$ is the position of the arrest front and $\xi = 1$ is a position equivalent to the posterior end of the notochord) at time $t$ is given [7] by:

$$\phi(\xi,t) = \Omega t + \gamma(1-\eta)^{-1}\left\{(1-\eta^2) - \rho^{-1}\xi\left[e^{-1/\rho} - \eta e^{-1/\rho\eta}\right] + \eta^2 e^{-\xi/\rho\eta} - e^{-\xi/\rho}\right\} + \phi_0, \qquad (S15)$$

where $\Omega$ is the collective frequency associated to the collective period, $\Omega = 2\pi/T_C$, and

$$\begin{aligned}\gamma &= \frac{\omega_A}{1-e^{-L/\sigma}} \cdot \frac{1+2\pi\varepsilon/\omega_A}{1+\varepsilon\tau} \cdot \frac{\sigma}{v}, \\ \eta &= \frac{\varepsilon}{2\sigma v} \cdot \frac{1+2\pi\varepsilon/\omega_A}{1+\varepsilon\tau}, \\ \rho &= \frac{\sigma}{L},\end{aligned} \qquad (S16)$$

with $\omega_A$ the frequency associated to the autonomous period of the oscillators, $\omega_A = 2\pi/T_A$, $L$ the length of the PSM from the arrest front to the posterior end of the notochord, $\sigma$ the decay length of the frequency profile along the PSM, $\varepsilon$ the coupling strength, $\tau$ the time delay in the coupling, and $v$ the velocity of the arrest front. $\phi_0$ is an arbitrary constant phase: in the movie it has been chosen such that the initial phase in the anterior border of the wildtype and the system with disrupted coupling are the same. The values of the independent parameters are given in the legend to Movie S1. We include a tailbud region at the posterior part of the PSM with homogeneous phase.

As the arrest front moves posteriorly (from left to right in the movie) the phase is arrested. The sine of the phase at a given point when the arrest front went through it is shown in grey scale in the left part of the movie: it is an illustration of the segmental pattern left behind by the oscillations in the PSM.

The frames comprising the movie were produced as ppm files from a C program encoding the formula for $\phi(\xi,t)$. The individual ppm files were assembled into an animated gif using the ImageMagick software package, and are available upon request.

### 2.2.9 Numerical simulations of the delayed coupling theory

For the simulations of the delayed coupling theory in main text Fig. 6, and in Figures S5 and S6, we used a hexagonal two-dimensional lattice that has the same average shape and size as the experimental conditions, (Fig. S10). In this hexagonal lattice the number of neighbors in the bulk is $N_i$=6, as illustrated by the red cell in Figure S10. In the boundary and corners, $N_i$ is smaller. We solved the DCT equations using parameters that we have estimated in this work for the wildtype zebrafish. We measured a cell size of 7 pixels, and used this as the size of single cells in the simulated patterns. Other parameters are the decay length of the frequency profile [7] $\sigma = 27$ cell diameters, the autonomous frequency of oscillators at the posterior boundary $\omega_A = 0.2205$ min$^{-1}$, the coupling strength $\varepsilon = 0.07$ min$^{-1}$, the coupling delay $\tau = 20.75$ min, and the arrest front velocity $v = 0.249$ cell diameters/min. Fluctuations were introduced with a noise term of strength $\eta = 0.02$ min$^{-1}$. From these simulations we obtained an average autocorrelation function describing noisy gene expression patterns (Experimental Procedures, and Supplemental Experimental Procedures 2.2.10) in excellent agreement with that obtained from the oscillatory expression pattern of *deltaC* in wildtype embryos (Fig. S5A; main text Fig. 6A, B).

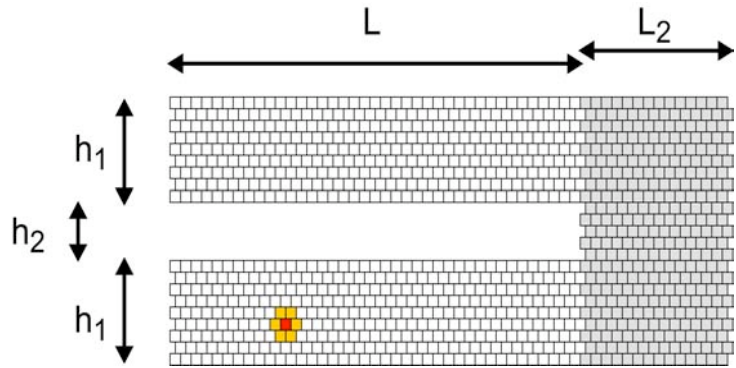

**Figure S10. Geometry of the hexagonal lattice used in the numerical simulations of the DCT in this work.** Spatial parameters are $L = 39$ cell diameters, $L_2 = 14$ cell diameters, $h_1 = 9$ cell diameters, $h_2 = 5$ cell diameters. The highlighted red cell has six yellow neighbors in the hexagonal lattice.

### 2.2.10 Autocorrelation function

Gene expression patterns are widely used as a manifestation of the state of the segmentation clock during somitogenesis. Phenotypes can be characterized by differing gene expression patterns, and alterations of such patterns from the wildtype pattern provide clues about the processes altered in the clock under different experimental conditions. While strong phenotypes display strong deviations from the wildtype pattern, subtle differences in the patterns cannot be reliably appraised without a quantitative analysis tool that allows the comparison of similar patterns in an unbiased way.

Furthermore, quantitative comparison of experimental conditions with theory calls for a quantitative analysis method regardless of the phenotypic strength.

Here we introduce a spatial autocorrelation function to measure disorder in gene expression patterns under different experimental conditions. We define the autocorrelation as

$$C(\delta) = \langle I(x) I(x+\delta) \rangle , \quad \text{(main text Eq. 6)} \quad \text{(S17)}$$

where $I(x)$ denotes the observed fluorescence intensity of cyclic gene expression at position $x$ measured along the antero-posterior axial direction as indicated in Fig. 6A in the main text, and the brackets denote an average over space along the same axial direction followed by an average across the lateral direction, as described below. This is a function of the distance $\delta$ between two points in the pattern, as indicated in Fig. 6A. Peaks of the correlation function occur at distances where the intensity of the pattern is similar.

We have also tried other definitions of the autocorrelation, *e.g.* subtracting the product of the averages $\langle I(x) \rangle \langle I(x+\delta) \rangle$ and normalizing by the variance $\langle I(x)^2 \rangle - \langle I(x) \rangle^2$. While some of these seemed at first sight a natural definition allowing for direct comparison of different datasets, we found that the interpretation of the curves was less straightforward because such definitions are very sensitive to the length of the box used to perform the analysis. This is because the largest characteristic length scale of the patterns that we study, the wavelength of the cyclic gene expression patterns, is similar to the length of the box where the autocorrelation function is computed. It is an unavoidable constraint that the pattern wavelength is of the same order as the system length. In the computation of single point averages such as $\langle I(x) \rangle$, $\langle I(x)^2 \rangle$ and $\langle I(x)^2 \rangle - \langle I(x) \rangle^2$, it makes a difference whether the length of the box is slightly smaller or slightly larger than the wavelength. Therefore, we recommend to leave them out of the definition of the autocorrelation function. Summarizing, of all the variants we tried, the definition (S17) is most robust and yields results which are simplest to interpret, and for this reason we use it in this work.

We used the same protocol to calculate the autocorrelations in experiments and in numerical simulations of the theory. In the simulations, we define the intensity as $I(x) = \sin(\phi(x))$, where $\phi(x)$ is the phase of oscillators at position $x$. We cropped a box from the anterior-most region of the PSM, where the most striking features of the expression patterns of wildtype, mutant, and also the Mib over-expression experiments are observed. We have computed the autocorrelation function for different box sizes, and in the posterior region. Longer boxes still showed the features we want to capture, but they were diluted because the correlations at the posterior introduce different features. The autocorrelation showed no difference in the posterior PSM of wildtype fish and Mib over-expression experiments.

Throughout this work we used a box of 50 x 120 pixels, with a resolution of 1.242 μm/pixel in the experiments (6B,D in main text). Images were taken in 8 bit grey scale format and converted to the ppm file format. We computed the autocorrelations of each sample with a custom computer code according to

$$C_i(\delta) = \frac{1}{50} \frac{1}{120} \sum_{x=1}^{120-\delta} \sum_{y=1}^{50} I_i(x,y) I_i(x+\delta,y) \,, \tag{S18}$$

where $C_i(\delta)$ is the autocorrelation of sample $i$, and $I_i(x,y)$ is the intensity of the pixel situated at $(x,y)$. After computing the autocorrelation for every single pattern, we averaged over the whole dataset to obtain the average autocorrelation

$$C(\delta) = \frac{1}{N} \sum_{i=1}^{N} C_i(\delta) \,. \tag{S19}$$

Because different experiments can have different background intensity levels and fluctuations, autocorrelations can have different absolute values. Here, whenever we compared the autocorrelation functions of different experimental conditions, we plotted them without changing the offset or scale, and displayed them in a box with the same range of values. When we compared to numerical simulations of the theory, we fitted the theory as described below by adjusting the offset and scale of the autocorrelations of simulations.

For the simulations, we used a two dimensional geometry that has the same average size as the experimental conditions. We measured a cell size of 7 pixels in experiments, and used this as the size of single cells in the simulated patterns. Because the phase is uniform inside each simulated cell, the theory cannot describe fluctuations occurring in experiments below the single cell length-scale from subcellular structure and organization. Furthermore, correlations at length-scales of about one cell diameter reflect artifacts of the regular lattice we use in the simulations, and thus we do not expect the simulations to provide a good fit to experimental data at these short length-scales either.

Our theory rests on a simplification of the individual autonomous oscillators, which are described by a phase variable. Amplitude is not accounted for in the theory, so amplitude effects and amplitude fluctuations occurring in the experiments cannot be described by the theory as it stands. In other words, although in the simulation two cells in the same phase will always have the same intensity, in the experiments, two cells with the same phase may differ in amplitude due to molecular fluctuations resulting in a different intensity. For this reason, when we compare the autocorrelations of theory and experiments, we fit the autocorrelations $C_{dct}(\delta)$ of simulations to the experimental autocorrelations $C_{exp}(\delta)$ using a linear function $F(\delta) = aC_{dct}(\delta) + b$, which introduces an offset $b$ and a scaling factor $a$. The offset accounts for background effects that depend on amplitude, and the scaling for differences in amplitude fluctuations that cannot be described by a theory based on phase oscillators. Thus, the offset and the scaling factor can be considered as fitting parameters.

In summary, using parameters of the delayed coupling theory determined from the experimental work in this paper, we simulate the spatial patterns of oscillating gene expression for various conditions. Their average autocorrelation functions are predictions about the experimental cases. From the experimentally observed patterns of oscillating gene expression in different conditions, we likewise calculate average autocorrelation functions and compare them to the predictions. In main text Figure 6C,E, the autocorrelation functions of experiments are plotted without any scaling or offset. The autocorrelation of numerical simulations using wildtype zebrafish parameters (Fig. 6C) is fit to the experiment by a scaling factor $a = 0.135$ and offset $b = 1068$, giving a coefficient of determination value of $R^2=0.96$. The Mib over-expression experiment in (Fig. 6D,E) is fit using a reduced value of the delay $\tau = 16.93$ min, with the same scaling factor obtained for the wildtype and an offset $b = -150$, yielding $R^2=0.77$. Error bars in Figure 6C,E are the standard error of the mean, *i.e.* the square root of the variance divided by the square root of the number of samples.

## *3. Supplemental References*

1. Oxtoby, E., and Jowett, T. (1993). Cloning of the zebrafish krox-20 gene (krx-20) and its expression during hindbrain development. Nucleic Acids Res *21*, 1087-1095.
2. Griffin, K.J., Amacher, S.L., Kimmel, C.B., and Kimelman, D. (1998). Molecular identification of spadetail: regulation of zebrafish trunk and tail mesoderm formation by T-box genes. Development *125*, 3379-3388.
3. Detrich, H.W., 3rd, Kieran, M.W., Chan, F.Y., Barone, L.M., Yee, K., Rundstadler, J.A., Pratt, S., Ransom, D., and Zon, L.I. (1995). Intraembryonic hematopoietic cell migration during vertebrate development. Proc Natl Acad Sci U S A *92*, 10713-10717.
4. Jiang, Y.J., Brand, M., Heisenberg, C.P., Beuchle, D., Furutani-Seiki, M., Kelsh, R.N., Warga, R.M., Granato, M., Haffter, P., Hammerschmidt, M., et al. (1996). Mutations affecting neurogenesis and brain morphology in the zebrafish, Danio rerio. Development *123*, 205-216.
5. van Eeden, F.J., Granato, M., Schach, U., Brand, M., Furutani-Seiki, M., Haffter, P., Hammerschmidt, M., Heisenberg, C.P., Jiang, Y.J., Kane, D.A., et al. (1996). Mutations affecting somite formation and patterning in the zebrafish, Danio rerio. Development *123*, 153-164.
6. Cooke, J., and Zeeman, E.C. (1976). A clock and wavefront model for control of the number of repeated structures during animal morphogenesis. J Theor Biol *58*, 455-476.
7. Morelli, L., Ares, S., Herrgen, L., Schroeter, C., Juelicher, F., and Oates, A. (2009). Delayed Coupling Theory of Vertebrate Segmentation. HFSP J *1*, 85.
8. Herrgen, L., Schroter, C., Bajard, L., and Oates, A.C. (2009). Multiple embryo time-lapse imaging of zebrafish development. Methods Mol Biol *546*, 243-254.
9. Schroter, C., Herrgen, L., Cardona, A., Brouhard, G.J., Feldman, B., and Oates, A.C. (2008). Dynamics of zebrafish somitogenesis. Dev Dyn *237*, 545-553.
10. Dequeant, M.L., and Pourquie, O. (2008). Segmental patterning of the vertebrate embryonic axis. Nat Rev Genet *9*, 370-382.
11. Lewis, J. (2003). Autoinhibition with transcriptional delay: a simple mechanism for the zebrafish somitogenesis oscillator. Curr Biol *13*, 1398-1408.
12. Tiedemann, H.B., Schneltzer, E., Zeiser, S., Rubio-Aliaga, I., Wurst, W., Beckers, J., Przemeck, G.K., and Hrabe de Angelis, M. (2007). Cell-based simulation of dynamic expression patterns in the presomitic mesoderm. J Theor Biol *248*, 120-129.
13. Westerfield, M. (2000). The zebrafish book. A guide for the laboratory use of zebrafish (Danio rerio). 4th Edition, (University of Oregon press).
14. Kimmel, C.B., Ballard, W.W., Kimmel, S.R., Ullmann, B., and Schilling, T.F. (1995). Stages of embryonic development of the zebrafish. Dev Dyn *203*, 253-310.
15. Holley, S.A., Geisler, R., and Nusslein-Volhard, C. (2000). Control of her1 expression during zebrafish somitogenesis by a delta-dependent oscillator and an independent wave-front activity. Genes Dev *14*, 1678-1690.
16. Julich, D., Hwee Lim, C., Round, J., Nicolaije, C., Schroeder, J., Davies, A., Geisler, R., Lewis, J., Jiang, Y.J., and Holley, S.A. (2005). beamter/deltaC and the role of Notch ligands in the zebrafish somite segmentation, hindbrain neurogenesis and hypochord differentiation. Dev Biol *286*, 391-404.
17. Holley, S.A., Julich, D., Rauch, G.J., Geisler, R., and Nusslein-Volhard, C. (2002). her1 and the notch pathway function within the oscillator mechanism that regulates zebrafish somitogenesis. Development *129*, 1175-1183.
18. Itoh, M., Kim, C.H., Palardy, G., Oda, T., Jiang, Y.J., Maust, D., Yeo, S.Y., Lorick, K., Wright, G.J., Ariza-McNaughton, L., et al. (2003). Mind bomb is a ubiquitin ligase that is essential for efficient activation of Notch signaling by Delta. Dev Cell *4*, 67-82.
19. Brand, M., Heisenberg, C.P., Jiang, Y.J., Beuchle, D., Lun, K., Furutani-Seiki, M., Granato, M., Haffter, P., Hammerschmidt, M., Kane, D.A., et al. (1996). Mutations in zebrafish genes affecting the formation of the boundary between midbrain and hindbrain. Development *123*, 179-190.
20. Heisenberg, C.P., Brand, M., Jiang, Y.J., Warga, R.M., Beuchle, D., van Eeden, F.J., Furutani-Seiki, M., Granato, M., Haffter, P., Hammerschmidt, M., et al. (1996). Genes involved in forebrain development in the zebrafish, Danio rerio. Development *123*, 191-203.
21. Grandel, H., Lun, K., Rauch, G.J., Rhinn, M., Piotrowski, T., Houart, C., Sordino, P., Kuchler, A.M., Schulte-Merker, S., Geisler, R., et al. (2002). Retinoic acid signalling in the zebrafish embryo is necessary during pre-segmentation stages to pattern the anterior-posterior axis of the CNS and to induce a pectoral fin bud. Development *129*, 2851-2865.
22. Geling, A., Steiner, H., Willem, M., Bally-Cuif, L., and Haass, C. (2002). A gamma-secretase inhibitor blocks Notch signaling in vivo and causes a severe neurogenic phenotype in zebrafish. EMBO Rep *3*, 688-694.
23. Riedel-Kruse, I.H., Muller, C., and Oates, A.C. (2007). Synchrony dynamics during initiation, failure, and rescue of the segmentation clock. Science *317*, 1911-1915.
24. Oates, A.C., and Ho, R.K. (2002). Hairy/E(spl)-related (Her) genes are central components of the segmentation oscillator and display redundancy with the Delta/Notch signaling pathway in the formation of anterior segmental boundaries in the zebrafish. Development *129*, 2929-2946.
25. Korzh, V., Edlund, T., and Thor, S. (1993). Zebrafish primary neurons initiate expression of the LIM homeodomain protein Isl-1 at the end of gastrulation. Development *118*, 417-425.
26. Sawada, A., Fritz, A., Jiang, Y.J., Yamamoto, A., Yamasu, K., Kuroiwa, A., Saga, Y., and Takeda, H. (2000). Zebrafish Mesp family genes, mesp-a and mesp-b are segmentally expressed in the presomitic mesoderm, and Mesp-b confers the anterior identity to the developing somites. Development *127*, 1691-1702.
27. Giudicelli, F., Ozbudak, E.M., Wright, G.J., and Lewis, J. (2007). Setting the tempo in development: an investigation of the zebrafish somite clock mechanism. PLoS Biol *5*, e150.
28. Markovsky, I., and Van Huffel, S. (2007). Overview of total least squares methods. Signal Process *87*, 2283.
29. Newman, M., and Barkema, G. (1999). Monte Carlo Methods in Statistical Physics, (Oxford: Oxford University Press).
30. Earl, M.G., and Strogatz, S.H. (2003). Synchronization in oscillator networks with delayed coupling: a stability criterion. Phys Rev E Stat Nonlin Soft Matter Phys *67*, 036204.
31. Jiang, Y.J., Aerne, B.L., Smithers, L., Haddon, C., Ish-Horowicz, D., and Lewis, J. (2000). Notch signalling and the synchronization of the somite segmentation clock. Nature *408*, 475-479.
32. Horikawa, K., Ishimatsu, K., Yoshimoto, E., Kondo, S., and Takeda, H. (2006). Noise-resistant and synchronized oscillation of the segmentation clock. Nature *441*, 719-723.
33. Ilagen, M., and Kopan, R. (2007). SanpShot: Notch signaling pathway. Cell *128*, 1246.e1241.
34. Weinmaster, G., and Kintner, C. (2003). Modulation of notch signaling during somitogenesis. Annu Rev Cell Dev Biol *19*, 367-395.
35. Rida, P.C., Le Minh, N., and Jiang, Y.J. (2004). A Notch feeling of somite segmentation and beyond. Dev Biol *265*, 2-22.
36. Chen, W., and Casey Corliss, D. (2004). Three modules of zebrafish Mind bomb work cooperatively to promote Delta ubiquitination and endocytosis. Dev Biol *267*, 361-373.
37. Le Borgne, R., Bardin, A., and Schweisguth, F. (2005). The roles of receptor and ligand endocytosis in regulating Notch signaling. Development *132*, 1751-1762.
38. Wang, W., and Struhl, G. (2004). Drosophila Epsin mediates a select endocytic pathway that DSL ligands must enter to activate Notch. Development *131*, 5367-5380.

39. Heuss, S.F., Ndiaye-Lobry, D., Six, E.M., Israel, A., and Logeat, F. (2008). The intracellular region of Notch ligands Dll1 and Dll3 regulates their trafficking and signaling activity. Proc Natl Acad Sci U S A *105*, 11212-11217.
40. Zhang, C., Li, Q., and Jiang, Y.J. (2007). Zebrafish Mib and Mib2 are mutual E3 ubiquitin ligases with common and specific delta substrates. J Mol Biol *366*, 1115-1128.
41. Zhang, C., Li, Q., Lim, C.H., Qiu, X., and Jiang, Y.J. (2007). The characterization of zebrafish antimorphic mib alleles reveals that Mib and Mind bomb-2 (Mib2) function redundantly. Dev Biol *305*, 14-27.
42. Momiji, H., and Monk, N.A. (2009). Oscillatory Notch-pathway activity in a delay model of neuronal differentiation. Phys Rev E Stat Nonlin Soft Matter Phys *80*, 021930.
43. Izhikevic, E.M. (1998). Phase models with explicit time delays. Phys Rev E Stat Nonlin Soft Matter Phys *58*, 905908.
44. Shimojo, H., Ohtsuka, T., and Kageyama, R. (2008). Oscillations in notch signaling regulate maintenance of neural progenitors. Neuron *58*, 52-64.